\documentclass{article}
\pdfoutput = 1

\usepackage{epsfig} 
\usepackage{amsmath,amsthm} 
\usepackage{amssymb}  
\newtheorem{thm}{Theorem}[section]
\newtheorem{lemma}{Lemma}[section]
\newtheorem{defi}{Definition}[section]
\newcommand{\beq}{\begin{equation}}
\newcommand{\eeq}{\end{equation}}

\title{\LARGE \bf
Lattices for Distributed Source Coding: Jointly Gaussian Sources and
Reconstruction of a Linear Function }

\author{Dinesh Krithivasan and S. Sandeep Pradhan\thanks{This work was
    supported by NSF grant (CAREER) CCF-0448115.}, \\
Department of Electrical Engineering and Computer Science, \\
University of Michigan, Ann Arbor, MI 48109, USA \\
email: {\tt\small dineshk@umich.edu, pradhanv@eecs.umich.edu}}

\date{}
\begin{document}
\bibliographystyle{ieeetr}

\maketitle
\thispagestyle{empty}
\pagestyle{plain}

\begin{abstract}
Consider a pair of correlated Gaussian sources $(X_1,X_2)$. Two
separate encoders observe the two components and communicate
compressed versions of their observations to a common decoder. The
decoder is interested in reconstructing a linear combination of
$X_1$ and $X_2$ to within a mean-square distortion of $D$. We obtain
an inner bound to the optimal rate-distortion region for this
problem. A portion of this inner bound is achieved by a scheme that
reconstructs the linear function directly rather than reconstructing
the individual components $X_1$ and $X_2$ first. This results in a
better rate region for certain parameter values. Our coding scheme
relies on lattice coding techniques in contrast to more prevalent
random coding arguments used to demonstrate achievable rate regions
in information theory. We then consider the case of linear
reconstruction of $K$ sources and provide an inner bound to the
optimal rate-distortion region. Some parts of the inner bound are
achieved using the following coding structure: lattice vector
quantization followed by ``correlated'' lattice-structured binning.
\end{abstract}


\section{Introduction} \label{sec:introduction}

Since its inception in 1973 by Slepian and Wolf, the problem of
distributed source coding has been a source of inspiration for
information/communication/data-compression  theory community because
of its formidable nature (in its full generality) and its wide scope
of practical applications.  In this problem, a collection of $K$
correlated information sources, with $i$th source having an alphabet
${\mathcal X}_i$, is observed separately by $K$ encoders. Each
encoder maps its observations into a finite-valued set. The indices
from these sets are transmitted over $K$ noiseless but
rate-constrained channels to a joint decoder. The decoder is
interested in obtaining $L$ reconstructions with $L$ fidelity
criteria (one for each). The $i$th reconstruction has an alphabet
$\hat{\mathcal X}_i$, and the $i$th fidelity criterion is a mapping
from the product of alphabets of a subset of the sources and
$\hat{\mathcal X}_i$ to the set of nonnegative real numbers. The
goal is to find a computable performance limit for this
communication problem. The performance limit, also referred to as
the optimal rate-distortion region, is expressed as the set of all
$(K+L)$-tuples of  rates of the $K$ indices transmitted by the
encoders and distortions of the $L$ reconstructions of the decoder
that can be achieved in the usual Shannon sense.

Toward this goal, progress has been made in a number of directions.
In the following we restrict our attention to the case of the
collection of stationary memoryless sources. In \cite{slepian73}, a
solution to the problem was given for the case when the decoder
wishes to reconstruct all the sources losslessly. In
\cite{wyner75,ahlswede-korner}, the case of lossless
``one-help-one'' problem was resolved. Here the decoder wishes to
reconstruct only one of the sources\footnote{The source which does
not enter into any of the fidelity criteria is referred to as a
helper. When the rate at which the helper is transmitted is greater
than its entropy, the helper is also referred to as side
information.} losslessly ($K=L+1=2)$. In \cite{wynerziv}, the case
of lossy ``one-help-one'' problem was resolved for the case when the
rate of the helper is greater than its entropy (also referred to as
the Wyner-Ziv problem). In \cite{berger77,bergertung}, an inner
bound, and an outer bound (also known as the Berger-Tung inner and
outer bounds respectively) to the performance limit are given for
the case where (a) $K=L=2$ and (b) the fidelity criterion of each
source does not depend on the other source (also referred to as
independent fidelity criteria). In \cite{berger-housewright}, an
inner bound to the performance limit is given for the case of lossy
``one-help-one'' problem. In \cite{gelfandpinsker}, an inner bound
to the performance limit is given for the case when the decoder
wishes to reconstruct a function of $K$ sources losslessly. It was
also shown that this is optimal for the case when the sources are
conditionally independent given the function. In
\cite{kornermarton}, the performance limit is given for
reconstructing losslessly the modulo-$2$ sum of two binary
correlated sources, and was shown to be tight for the symmetric
case. This has been extended to several cases in \cite{csiszar} (see
Problem 23 on page 400) and \cite{han-kobayashi}. An improved inner
bound was provided for this case in  \cite{ahlswede-han}. The key
point to note is that the performance limits given in
\cite{kornermarton,han-kobayashi,ahlswede-han} are outside the inner
bound given in \cite{gelfandpinsker}. In \cite{berger-yeung}, the
performance limit is given for the case where (a) $K=L=2$, (b) one
of the sources is reconstructed losslessly and the other with a
independent fidelity criterion. In \cite{viswanathan-berger-new}
(also see
\cite{han79,viswanathan-berger-old,viswanath,yamamoto-itoh,flynn-gray,zamir-berger}),
an inner bound to the performance limit of the CEO problem
\footnote{This is a variant of the general distributed source coding
problem mentioned above. This is closely related to another class of
distributed source coding problems known as remote source coding
problems. Here the encoders observe a noisy version of the sources.
However it can be shown using the techniques of
\cite{dobrushin,witsenhausen} that the remote source coding problems
are equivalent to a class of general distributed source coding
problems mentioned above.} was given. This problem for the quadratic
Gaussian case essentially boils down to reconstructing a certain
linear function of the sources with mean squared error fidelity
criterion. It was shown that this inner bound is tight for some
cases in \cite{oohama98,prabhakaran-ramchandran-tse}. For the vector
Gaussian CEO problem, inner and outer bounds were derived in
\cite{oohama_allerton05,oohama_ita06}. These bounds were shown to be
tight under some conditions. In \cite{orlitsky-roche}, the
performance limit is given for the case of lossless reconstruction
of a function of two sources with the rate of one of the sources
being greater than or equal to its entropy. The lossy version is
addressed in \cite{yamamoto,feng-effros-savari}. Regarding the
Berger-Tung inner bound, it was shown that this is tight for (a) the
high-resolution case with independent fidelity criteria in
\cite{zamir-berger},  (b) the jointly Gaussian case  $K=2$, $L=1$
and independent squared error fidelity criterion in \cite{oohama97},
and  (c) the jointly Gaussian case with $K=2,L=2$ and independent
squared error criteria in \cite{wagner-tavildar-viswanath-old}. In
\cite{wagner-tavildar-viswanath-old}, it was also shown that a
Berger-Tung based coding scheme is optimal for the case of
reconstruction of certain linear functions of two jointly Gaussian
sources with squared error criterion. A general outer bound to the
performance limit of the general distributed source coding problem
was given in \cite{wagner-anantharam}. In \cite{gastpar}, the
performance limit was given for the lossy ``one-help-many'' problem
with independent fidelity criteria and the sources being
conditionally independent given the helper which is transmitted at a
rate greater than its entropy. In \cite{oohama05}, the performance
limit was given for the quadratic jointly Gaussian lossy
``many-help-one'' problem with the condition that the helpers are
conditionally independent given the source. In
\cite{wagner-tavildar-viswanath-new}, the performance limits were
obtained for the case of quadratic Gaussian ``many-help-one''
problem where the sources satisfy a ``tree-structure''. In
\cite{jana-blahut}, the performance limit is given for the case
where one of the sources needs to be reconstructed with an
independent fidelity criterion and the rest of the sources need to
be reconstructed losslessly. In \cite{jana1}, infinite order
descriptions were provided for the performance limits of the general
case of two terminal source coding problem ($K=2$) with independent
distortion criteria. This was extended to the case of more than two
sources in \cite{jana2}.

With regard to above set of results, we would like to make the
following observations. (a) Most of the above approaches, except
that of \cite{kornermarton} and its extensions in \cite{csiszar,
han-kobayashi, ahlswede-han}, use random vector quantization
followed by independent random binning  (see Chapter 14 of
\cite{cover-thomas})  of the quantizer indices.  (b) The four
exceptions, which consider only lossless source coding problems,
deviate from this norm, and instead use structured random binning
based on linear codes on finite fields. Further, the binning
operation of the quantizers of the sources are ``correlated''. This
incorporation of structure in binning appears to give improvements
in the rates especially for those cases that involve reconstruction
of a function of the sources. Moreover, it is still not known
whether it is possible to approach this performance without the
structured codes. (c) For some distributed source coding problems
(that belong to the first category), whose performance limits were
derived using random coding and random binning, it is well-known
that these limits can also be approached using structured codes. For
example structured codes were considered for (a) the Slepian-Wolf
problem in \cite{csiszar82},  (b)  the Wyner-Ziv problem for the
binary case with Hamming distortion and for the quadratic Gaussian
case in \cite{zamirmulti}, (c) the Berger-Tung inner bound for the
two terminal quadratic Gaussian problem with independent fidelity
criteria in \cite{zamirmulti} and (d) high-resolution distributed
source coding problem with independent fidelity criteria in
\cite{zamir-berger}.

With this as a motivation, in this paper we consider a lossy
distributed source coding problem with $K$ jointly Gaussian sources
with one reconstruction, i.e., $L=1$.  The fidelity criterion has
the additional structure that is given by the following. The decoder
wishes to reconstruct a linear function of the sources with squared
error as the fidelity criterion.  We consider a coding scheme with
the following structure: sources are quantized  using structured
vector quantizers followed by ``correlated'' structured binning.
That is, the binning operations of the quantizers of the sources are
not performed ``independently''. The structure used in this process
is given by lattice codes. We provide an inner bound to the optimal
rate-distortion region. We show that the proposed inner bound is
better for certain parameter values than an inner bound that can be
obtained by using a coding scheme that uses random vector quantizers
following by independent random binning. For this purpose we use the
machinery developed by
\cite{zamir-feder,zamirlqn,zamirmulti,zamirsnr,zamirgood} for the
Wyner-Ziv problem in the quadratic Gaussian case.

We also believe that the proposed scheme can be used as a building
block to provide an inner bound to the optimal rate-distortion
region for the case when the decoder wishes to reconstruct all the
sources with independent squared error fidelity criterion. This will
be addressed in our future work. The rest of the paper is organized
as follows. Rather than giving the main result for the most general
case first and then considering special cases, we first consider the
case of two sources and obtain the result and then provide the
result for the general case. In Section \ref{sec:latticesall}, we
give a concise overview of the asymptotic properties of
high-dimensional lattices that are known in the literature and we
use these properties  in the rest of the paper. In Section
\ref{sec:defandresult}, we define the problem formally for the case
of two sources and present an inner bound to the optimal
rate-distortion region given by a coding structure involving
structured quantizers followed by ``correlated'' structured binning.
Further, we also present another inner bound achieved by a scheme
that first obtains a lossy reconstruction of the sources, and then
obtains a reconstruction of the linear function. The latter scheme
is based on the Berger-Tung inner bound. An overall achievable rate
region can be obtained by combining these two schemes. Then we
present our lattice based coding scheme and prove achievability of
the inner bound.  We also provide motivation and intuition about the
proposed  coding scheme in this section. In Section
\ref{sec:generalizations}, we consider a generalization of the
problem that involves reconstruction of a linear function of an
arbitrary finite number of sources. We also demonstrate how the
general solution simplifies in certain special cases. In Section
\ref{sec:comparison}, we provide a set of numerical results for the
two-source case that demonstrate the conditions under which the
lattice based scheme performs better than the Berger-Tung based
scheme.  We conclude with some comments in Section
\ref{sec:conclusions}.

A word about the notation used in this paper is in order. Let
$f(\cdot)$ be an arbitrary function that takes as input a scalar.
Then the function $f^n(\cdot)$ takes an $n$-length vector as input
and operates component-wise on the components of that vector. This
notation generalizes to functions of more than one variable as well.
Variables with superscript $n$ denote an $n$-length random vector
whose components are mutually independent. However, random vectors
whose components are not independent are denoted without the use of
the superscript. The dimension of such random vectors will be clear
from the context.

\section{Preliminaries on high-dimensional Lattices} \label{sec:latticesall}
\subsection{Overview of Lattice Codes} \label{subsec:lattices} Lattice
codes \cite{conway-sloane} play the same role in Euclidean space
that linear codes play in Hamming space. Introduction to lattices
and to coding schemes that employ lattice codes can be found in
\cite{zamirlqn,zamirmulti,zamirsnr,loeliger,kirac-vaidyanathan}.
Lattice codes have been used in other related multiterminal source
coding problems in the literature
\cite{vaishampayan-sloane-servetto, dayan-zamir,
goyal-kelner,diggavi-sloane-vaishampayan, ostergaard}. In the rest
of this section, we will briefly review some properties of lattice
codes that are relevant to our coding scheme. We start by defining
various quantities of interest associated with lattices. We use the
same notation as in \cite{zamirmulti} for these quantities.

An n-dimensional lattice $\Lambda$ is composed of all integer
combinations of the columns of an $n \times n$ matrix $G$ called the
generator matrix of the lattice.
\begin{align} \label{eq:latdef}
\Lambda &= \{l \in \mathbb{R}^n \, \colon \, l = G \cdot i \text{
for some } i \in \mathbb{Z}^n \}
\end{align}
Associated with every lattice $\Lambda$ is a natural quantizer
namely one that associates with every point in $\mathbb{R}^n$ its
nearest lattice point. This quantizer can be described by the
function
\begin{align} \label{eq:Qdef}
Q_{\Lambda}(x) &\triangleq l \in \Lambda \text{ where } \parallel
x-l \parallel \, \leq \, \parallel x - \hat{l} \parallel \quad
\text{for all } \hat{l} \in \Lambda.
\end{align}
The quantization error associated with the quantizer
$Q_{\Lambda}(\cdot)$ is defined by
\begin{align} \label{eq:moddef}
x \text{ mod } \Lambda &= x - Q_{\Lambda}(x).
\end{align}
The basic Voronoi region of a lattice $\Lambda$ is the set of all
points closer to the origin than to any other lattice point, i.e.,
\begin{align} \label{eq:voronoidef}
\mathcal{V}_0(\Lambda) &= \{ x \in \mathbb{R}^n \, : \, Q_{\Lambda}(x) = 0^n
\}
\end{align}
where $0^n$ is the origin of $\mathbb{R}^n$. The second moment of a
lattice $\Lambda$ is the expected value per dimension of the norm of
a random vector uniformly distributed over $\mathcal{V}_0(\Lambda)$ and is
given by
\begin{align} \label{eq:siglatdef}
\sigma^2(\Lambda) &= \frac{1}{n} \frac {\int_{\mathcal{V}_0(\Lambda)}
\parallel x \parallel^2 \text{dx}}{\int_{\mathcal{V}_0(\Lambda)}
\text{dx}}
\end{align}
Let the normalized second moment be give by \beq
\label{eq:latnorm2mom}
G(\Lambda)=\frac{\sigma^2(\Lambda)}{V^{2/n}(\Lambda)} \eeq where
$V(\Lambda)=\int_{\mathcal{V}_0(\Lambda)} \text{dx}$. When used as a
channel code over an unconstrained AWGN channel with noise $Z$
having variance $\sigma^2_Z$ \cite{Poltyrev}, let the probability of
decoding error be denoted by \beq P_e(\Lambda,\sigma_Z^2)=Pr(Z^n
\not \in {\mathcal V}_0) \eeq where $Z^n$ is the random noise vector
of length $n$.

The mod operation defined in equation (\ref{eq:moddef}) satisfies
the following useful distributive property.
\begin{align} \label{eq:latdistributive}
((x \text{ mod } \Lambda) + y ) \text{ mod } \Lambda &= (x+y) \text{
mod } \Lambda \quad \forall \, x,y .
\end{align}

It is known (see \cite{zamirlqn} \cite{zamirsnr}) that the
quantization error of a lattice quantizer $\Lambda$ can be assumed
to have a nearly uniform distribution over the fundamental Voronoi
region $\mathcal{V}_0$ of the quantizer. This assumption is
completely accurate in the case of subtractive dithered quantization
where a vector uniformly distributed over $\mathcal{V}_0$ (called
the dither) is added at the encoder before quantization and
subtracted at the decoder. It has been shown in \cite{zamirlqn} that
for an optimal lattice quantizer, this noise is wide-sense
stationary and white. Further, as the lattice dimension $n
\rightarrow \infty$, for optimal lattice quantizers, the
quantization noise approaches a white Gaussian noise process in the
Kullback-Leibler divergence sense.

Lattices have been studied extensively for efficient packing and
covering. A systematic study of lattice packings was initiated by
Minkowski in \cite{minkowski_1904}, where existence of good lattice
packings was shown. A formal study of lattice covering appears to
have been initiated by Kershner in \cite{kershner_1939}. See
\cite{rogers_book} for a thorough review of existence of efficient
lattice packings and coverings. Lattice codes have been employed in
the point-to-point setting for quantization of Gaussian sources with
squared error fidelity criterion and also in coding for the AWGN
channel with power constraint. In \cite{zamirmulti}, the existence
of high dimensional lattices that are ``good'' for quantization and
for coding is discussed. The criteria used therein to define
goodness are as follows:
\begin{itemize}
\item A sequence of lattices $\Lambda^{(n)}$ (indexed by the dimension
  $n$)  is said to be a good channel $\sigma_Z^2$-code sequence
if $\forall \epsilon>0$, there exists $N(\epsilon)$ such that
for all $n>N(\epsilon)$ the following conditions are satisfied:
\beq
 V(\Lambda^{(n)}) < 2^{n( \frac{1}{2} \log (2 \pi e \sigma_Z^2)+
  \epsilon)},
\eeq \beq P_e(\Lambda^{(n)},\sigma_Z^2) < 2^{-nE(\epsilon)} \eeq for
some $E(\epsilon)>0$.
\item  A sequence of lattices $\Lambda^{(n)}$ (indexed by the dimension
  $n$)  is said to be a good source $D$-code sequence
if $\forall \epsilon>0$, there exists $N(\epsilon)$ such that for
all $n>N(\epsilon)$ the following conditions are satisfied: \beq
\log ( 2 \pi e G(\Lambda^{(n)})) < \epsilon \eeq \beq
\sigma^2(\Lambda^{(n)})=D. \eeq
\end{itemize}

\subsection{Nested Lattice Codes} \label{subsec:nestedlattices}

For lossy coding problems involving side-information at the
encoder/decoder, it is natural to consider nested codes. Wyner
proposed an algebraic binning approach involving linear codes for
the Slepian-Wolf problem \cite{wyner74}. Adapting this scheme to the
case of lossy coding, nested codes for the Wyner-Ziv problem were
proposed in \cite{shamai-verdu-zamir}. We review the properties of
nested lattice codes briefly here. Further details can be found in
\cite{zamirmulti}.

A pair of $n$-dimensional lattices $(\Lambda_1,\Lambda_2)$ is
nested, i.e., $\Lambda_2 \subset \Lambda_1$, if their corresponding
generating matrices $G_1,G_2$ satisfy
\begin{align}
G_2 = G_1 \cdot J
\end{align}
where $J$ is an $n \times n$ integer matrix with determinant greater
than one. $\Lambda_1$ is referred to as the fine lattice while
$\Lambda_2$ is the coarse lattice. The points of the set
\begin{align}
\{ \Lambda_1 \text{ mod } \Lambda_2 \} \triangleq \{ \Lambda_1 \cap
\mathcal{V}_{0,2} \}
\end{align}
are called the coset leaders of $\Lambda_2$ relative to $\Lambda_1$.
The nesting ratio of this nested lattice is defined as
$\sqrt[n]{V_2/V_1}$ where $V_i = V(\Lambda_i)$ is the volume of the
Voronoi region of lattice $\Lambda_i$, $i = 1,2$.

In many applications of nested lattice codes, we require the
lattices involved to be a good source code and/or a good channel
code. We term a nested lattice $(\Lambda_1,\Lambda_2)$ good if (a)
the fine lattice $\Lambda_1$ is both a good $\delta_1$-source code
and a good $\delta_1$-channel code and (b) the coarse lattice
$\Lambda_2$ is both a good $\delta_2$-source code and a
$\delta_2$-channel code. For such a nested lattice code
$(\Lambda_1,\Lambda_2)$, the number of coset leaders of $\Lambda_2$
relative to $\Lambda_1$ is about $(\delta_2 / \delta_1)^{n/2}$. A
code employing the coset leaders as codewords would thus have a rate
of $\frac{1}{2} \log (\delta_2 / \delta_1)$. Equivalently, the rate
of such a code is the logarithm of the nesting ratio of the nested
lattice $(\Lambda_1,\Lambda_2)$.

The existence of good lattice codes and good nested lattice codes
(for various notions of goodness) has been studied in
\cite{zamirsnr, zamirgood} which use the random coding method of
\cite{hlawka, loeliger}. In \cite{zamirgood}, it was shown that
there exists lattices which are simultaneously good in both the
source and channel coding senses described above. In
\cite{zamirsnr}, the existence of nested lattices where the coarse
lattice is simultanously good as a source and channel code and the
fine lattice is a good channel code was proved.

\section{Distributed source coding for the two-source case} \label{sec:defandresult}
\subsection{Problem Statement and Main Result} \label{subsec:formaldef}

In this section we consider a distributed source coding problem for
the case of two sources $X_1$ and $X_2$. The function to be
reconstructed at the decoder is assumed to be the linear function $Z
\triangleq F(X_1,X_2) = X_1 - c X_2$ unless otherwise specified.
Consideration of this function is enough to infer the behavior of
any linear function $c_1X_1 + c_2X_2$ and has the advantage of fewer
variables. We consider the more general case of $F(X_1,\dots,X_K) =
\sum_{i=1}^{K} c_i X_i$ in Section \ref{sec:generalizations}.

We define the coding problem formally below. Consider a pair of
correlated jointly Gaussian sources $(X_1,X_2)$ with a given joint
distribution $p_{X_1 X_2}(x_1,x_2)$.  The source sequence
$(X_1^n,X_2^n)$ is independent over time and has the product
distribution $\prod_{i=1}^{n} p_{X_1 X_2}(x_{1i},x_{2i})$. Consider
the following average squared error as the fidelity criterion: $d :
\mathbb{R}^n \times \mathbb{R}^n \rightarrow \mathbb{R}^{+}$ given
by
\begin{align}
d(x^n,y^n) &= \frac{1}{n} \sum_{i=1}^{n} (x_i-y_i)^2.
\end{align}
\begin{defi}
Given such a jointly Gaussian distribution $p_{X_1 X_2}$ and a
distortion function $d(\cdot,\cdot)$ a transmission system  with
parameters $(n,\theta_1,\theta_2,\Delta)$ is defined as the set of
mappings
\begin{align}
f_i &: \mathbb{R}^n \rightarrow \{1,2,\dots,\theta_i \} \quad
\text{for }i=1,2 \\
g &: \{1,2,\dots,\theta_1\} \times \{1,2,\dots,\theta_2\}
\rightarrow \mathbb{R}^n
\end{align}
such that the following constraint is satisfied
\begin{align}
\mathbb{E} \left( d(F^n(X_1^n,X_2^n),g(f_1(X_1^n),f_2(X_2^n)))
\right) &\leq \Delta.
\end{align}
We say that a tuple $(R_1,R_2,D)$ is achievable if $\forall
\epsilon > 0$, $\exists$ for all sufficiently large $n$, a
transmission system with parameters $(n,\theta_1,\theta_2,\Delta)$
such that
\begin{align*}
\frac{1}{n} \log \theta_i &\leq R_i + \epsilon \quad \text{for }i
=1,2 \\
\Delta &\leq D + \epsilon.
\end{align*}
The performance limit is given by the optimal rate-distortion region
which is defined as the set of all achievable tuples $(R_1,R_2,D)$.
This problem is graphically illustrated in Fig. \ref{fig:blockdia}.
\begin{figure}[htp]
\centering
\includegraphics[width = 0.7\textwidth]{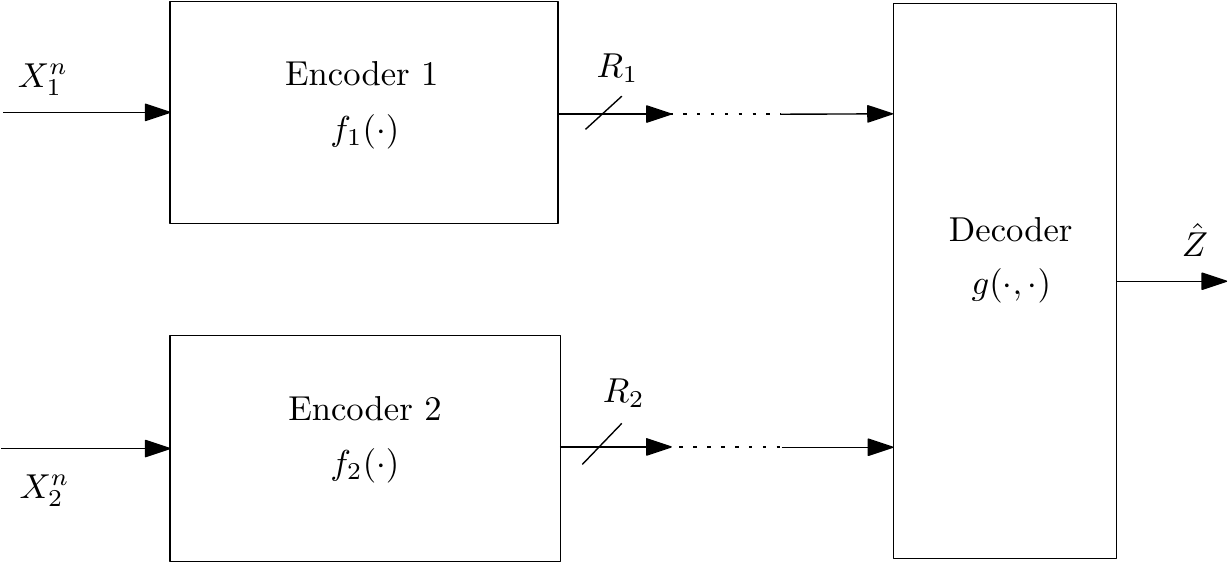}
\caption{Schematic Illustration of the problem} \label{fig:blockdia}
\end{figure}
\end{defi}
Without loss of generality, the sources can be assumed to have unit
variance and let the correlation coefficient $\rho > 0$. For the
rest of this section, these assumptions are made unless otherwise
stated.

One possible coding scheme for this problem would be the following.
The decoder reconstructs lossy versions $(W_1,W_2)$ of the sources
$(X_1,X_2)$ and uses the best estimate of $Z$ given $(W_1,W_2)$ as
the reconstruction $\hat{Z}$. The rate region for such a scheme can
be derived using the Berger-Tung inner bound
\cite{berger77,bergertung}. One of the main result in this paper is
to show that for certain parameter values, there exists a better
coding scheme that enables the decoder to reconstruct $\hat{Z}$
directly without resorting to reconstructions $(W_1,W_2)$. We
present the rate region of our scheme below.

\begin{thm} \label{thm:latticerate}
The set of all tuples of rates and distortion $(R_1,R_2,D)$ that
satisfy
\begin{align}
2^{-2R_1} + 2^{-2R_2} &\leq \left(\frac{\sigma_Z^2}{D} \right)^{-1}
\end{align}
are achievable. Here, $\sigma^2_Z = 1+c^2-2\rho c$ is the variance
of the function $Z$ to be reconstructed.
\end{thm}
\textbf{Proof:} See Section \ref{subsec:mainscheme}.

We also present another achievable rate region based on ideas
similar to the Berger-Tung coding scheme \cite{berger77}
\cite{bergertung}. From here on, we shall refer to this rate region
as the Berger-Tung based rate region and the scheme that achieves
this as the Berger-Tung based coding scheme.

\begin{thm} \label{thm:bergertungrate}
Let the region $\mathcal{RD}_{\text{in}}$ be defined as follows.
\begin{equation*}
\mathcal{RD}_{\text{in}} = \bigcup_{(q_1,q_2) \in \mathbb{R}^2_{+}}
\left\{ (R_1,R_2,D) \colon R_1 \geq \frac{1}{2} \log
\frac{(1+q_1)(1+q_2) - \rho^2}{q_1(1+q_2)},
\, \, R_2 \geq
\frac{1}{2} \log \frac{(1+q_1)(1+q_2)-\rho^2}{q_2(1+q_1)} \right.
\end{equation*}
\begin{equation} \label{eq:BTrateregion}
\left. R_1+R_2 \geq \frac{1}{2} \log
\frac{(1+q_1)(1+q_2)-\rho^2}{q_1 q_2}, \, \, D \geq \frac{q_1 \alpha
+ q_2 c^2 \alpha + q_1 q_2 \sigma_Z^2}{(1+q_1)(1+q_2)-\rho^2}
\right\}.
\end{equation}
where $\alpha \triangleq 1-\rho^2$ and $\Bbb{R}_{+}$ is the set of
positive reals. Then the rate distortion tuples $(R_1,R_2,D)$ which
belong to $\mathcal{RD}^{*}_{\text{in}}$ are achievable where $^*$
denotes convex closure.
\end{thm} \textbf{Proof:} Follows directly from the application of
Berger-Tung inner bound with the auxiliary random variables involved
being Gaussian.

In many distributed source coding problems involving jointly
Gaussian sources
(\cite{oohama98,prabhakaran-ramchandran-tse,wagner-tavildar-viswanath-old}),
the use of Gaussian auxiliary random variables results in the
optimal or largest known rate region. It was conjectured in
\cite{berger77,bergertung} that choosing the auxiliary random
variables to be Gaussian indeed results in the optimal rate
distortion region for the problem of reconstructing both sources
with independent distortion criteria. This was shown to be true in
\cite{wagner-tavildar-viswanath-old}. With this as motivation, we
have used Gaussian auxiliary random variables to derive an inner
bound to the performance limit of this problem based on the
Berger-Tung coding scheme.

We have the following lemma that gives the minimum sum rate of the
second approach which will be used in later sections for comparing
the performance limits given by the above two theorems.
\begin{lemma}
For a given distortion $D$, the minimum sum rate $R_{\text{sum}}
\triangleq R_1+R_2$ that lies in the region
$\mathcal{RD}_{\text{in}}^{*}$ of Theorem \ref{thm:bergertungrate}
is given by the lower convex envelope of the following region. \beq
\label{eq:BTratecommon} R_{\text{sum}} \geq \frac{1}{2} \log
\frac{4c(\alpha c-\rho D)}{D^2} \quad D \leq \min \left\{ \frac{2
\alpha c}{\rho+c}, \frac{2\alpha c^2}{1+\rho c} \right\} \eeq \beq
\label{eq:BTratecless1} R_{\text{sum}} \geq \frac{1}{2} \log \left(
\frac{(1 - \rho c)^2}{D-\alpha c^2} \right) \quad \sigma_Z^2 > D >
\frac{2 \alpha c^2}{1+\rho c}, \, c \leq 1 \eeq \beq
\label{eq:BTratecmore1} R_{\text{sum}} \geq \frac{1}{2} \log \left(
\frac{(c - \rho)^2}{D-\alpha} \right) \quad \sigma_Z^2 > D > \frac{2
\alpha c}{\rho+ c}, \,\, c > 1 \eeq \beq R_{\text{sum}} = 0 \quad D
\geq \sigma_Z^2 \eeq
\end{lemma}
\textbf{Proof:} This derivation is detailed in Appendix
\ref{sec:btderive}.

For certain values of $\rho$, $c$ and $D$, the sum-rate given by
Theorem \ref{thm:latticerate} is better than that given in Theorem
\ref{thm:bergertungrate}. This implies that each rate region
contains rate points which are not contained in the other. Thus, an
overall achievable rate region for the coding problem can be
obtained as the convex closure of the union of all rate distortion
tuples $(R_1,R_2,D)$ given in Theorems \ref{thm:latticerate} and
\ref{thm:bergertungrate}. A further comparison of the two schemes is
presented in Section \ref{sec:comparison}. Note that for $c<0$, it
has been shown in \cite{wagner-tavildar-viswanath-old} that the rate
region given in Theorem \ref{thm:bergertungrate} is tight.

\subsection{The Coding Scheme} \label{subsec:mainscheme} In this
section, we present a lattice based coding scheme for the problem of
reconstructing the above linear function of two jointly Gaussian
sources whose performance approaches the inner bound given in
Theorem \ref{thm:latticerate}. In what follows, a nested lattice
code is taken to mean a sequence of nested lattice codes indexed by
the lattice dimension $n$.

We will require nested lattice codes
$(\Lambda_{11},\Lambda_{12},\Lambda_2)$ where $\Lambda_2 \subset
\Lambda_{11}$ and $\Lambda_2 \subset \Lambda_{12}$. We need the fine
lattices $\Lambda_{11}$ and $\Lambda_{12}$ to be good source codes
(of appropriate second moment) and the coarse lattice $\Lambda_2$ to
be a good channel code. The proof of the existence of such nested
lattices is detailed in Appendix \ref{sec:nestedlatticeproof} where
we show the existence of a nested lattice $(\Lambda_{11},
\Lambda_{12}, \Lambda_2)$ such that $\Lambda_{11} \subset
\Lambda_{12} \subset \Lambda_2$ or $\Lambda_{12} \subset
\Lambda_{11} \subset \Lambda_2$ and all three lattices are good
source and channel codes simultaneously. The parameters of the
nested lattice are chosen to be
\begin{align}
\sigma^2(\Lambda_{11}) &= q_1 \label{eq:latsig1} \\
\sigma^2(\Lambda_{12}) &= \frac{D \sigma_Z^2}{\sigma_Z^2-D} - q_1.
\label{eq:latsig2} \\
\sigma^2(\Lambda_2) &= \frac{\sigma_Z^4}{\sigma_Z^2-D}
\end{align}
where $0 < q_1 < D\sigma_Z^2/(\sigma_Z^2-D)$. The coding problem is
non-trivial only for $D<\sigma^2_Z$ and in this range,
$D\sigma_Z^2/(\sigma_Z^2-D) < \sigma^2(\Lambda_2)$ and therefore
$\Lambda_2 \subset \Lambda_{11}$ and $\Lambda_2 \subset
\Lambda_{12}$ indeed. Note that the order of nesting between the
lattices $\Lambda_{11}$ and $\Lambda_{12}$ depends on whether $q_1 >
D\sigma_Z^2/2(\sigma_Z^2-D)$ or not. However, this is irrelevant for
the proof which only requires $\Lambda_2 \subset \Lambda_{11}$ and
$\Lambda_2 \subset \Lambda_{12}$.

\begin{figure*}[htp]
\centering
\includegraphics[width = \textwidth]{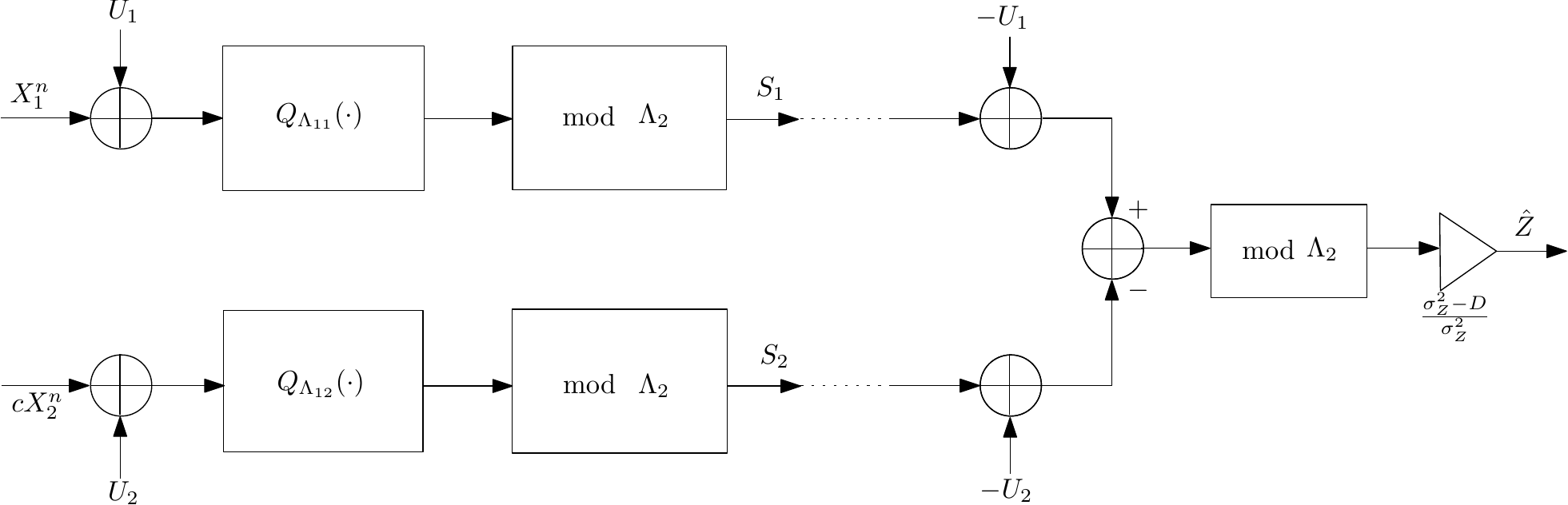}
\caption{Distributed coding using lattice codes to reconstruct $Z =
X_1 - c X_2$} \label{fig:fullrep}
\end{figure*}

Let $U_1$ and $U_2$ be random vectors (dithers) that are independent
of each other and of the source pair $(X_1,X_2)$. Let $U_i$ be
uniformly distributed over the basic Voronoi region
$\mathcal{V}_{0,1i}$ of the fine lattices $\Lambda_{1i}$ for
$i=1,2$. The decoder is assumed to share this randomness with the
encoders. The source encoders use these nested lattices to quantize
$X_1$ and $cX_2$ respectively according to equation
\begin{align}
S_1 &= \left( Q_{\Lambda_{11}}(X_1^n + U_1) \right) \text{ mod }
\Lambda_2 \label{eq:encdesc1},\\
S_2 &= \left( Q_{\Lambda_{12}}(cX_2^n + U_2) \right) \text{ mod }
\Lambda_2 \label{eq:encdesc2}.
\end{align}
Note that the second encoder scales the source $X_2$ before encoding
it. The decoder receives the indices $S_1$ and $S_2$ and
reconstructs
\begin{align} \label{eq:decdesc}
\hat{Z} &= \left( \frac{\sigma_Z^2-D}{\sigma_Z^2} \right) \left( [
(S_1 -U_1) - (S_2-U_2)] \text{ mod } \Lambda_2 \right).
\end{align}

This coding scheme is illustrated in Fig. \ref{fig:fullrep}. The
rates of the two encoders are given by
\begin{equation} \label{eq:R12user}
R_1 = \frac{1}{2} \log \frac{\sigma_Z^4}{q_1(\sigma_Z^2-D)}
\end{equation}
\begin{equation} \label{eq:R22user}
R_2 = \frac{1}{2} \log
\frac{\sigma_Z^4}{D\sigma_Z^2-q_1(\sigma_Z^2-D)}
\end{equation}
Clearly, for a fixed choice of $q_1$ all rates greater than those
given in equations (\ref{eq:R12user}) and (\ref{eq:R22user}) are
achievable. The union of all achievable rate-distortion tuples
$(R_1,R_2,D)$ over all choices of $q_1$ gives us an achievable
region. Eliminating $q_1$ between the two rate equations gives us
\beq \label{eq:Rsumproof} 2^{2R_2} \geq
\frac{1}{\frac{D}{\sigma_Z^2}-2^{-2R_1}} \eeq which is the rate
region claimed in Theorem \ref{thm:latticerate}. It remains to show
that this scheme indeed reconstructs the function $Z$ to within a
distortion $D$. We show this in the following.

\begin{figure}[htp]
\centering
\includegraphics[width = 0.6\textwidth]{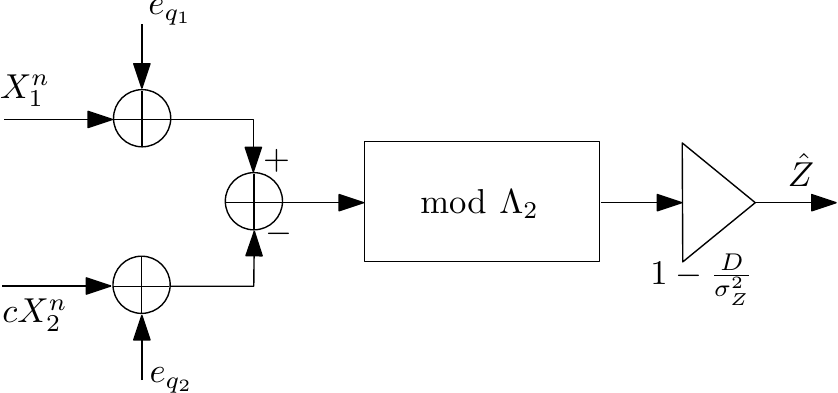}
\caption{Equivalent representation of Fig. \ref{fig:fullrep}}
\label{fig:eqrep}
\end{figure}

Using the distributive property of lattices described in equation
(\ref{eq:latdistributive}), we can reduce the coding scheme to a
simpler equivalent scheme by eliminating the first mod-$\Lambda_2$
operation in both the signal paths. This results in an equivalent
representation of the coding scheme as shown in Fig.
\ref{fig:eqrep}. The decoder can now be described by the equation
\begin{align}
\hat{Z} &= \left(\frac{\sigma_Z^2-D}{\sigma_Z^2} \right) ([(X_1^n +
e_{q_1}) - (cX_2^n + e_{q_2})] \text{ mod } \Lambda_2) \\ &= \left(
\frac{\sigma_Z^2-D}{\sigma_Z^2} \right) \left([Z^n+e_{q_1}-e_{q_2}]
\text{ mod } \Lambda_2 \right) \label{eq:decoderfinal}
\end{align}
where $e_{q_1}$ and $e_{q_2}$ are dithered lattice quantization
noises given by
\begin{align}
e_{q_1} &= Q_{\Lambda_{11}}(X_1^n+U_1) - (X_1^n+U_1), \\
e_{q_2} &= Q_{\Lambda_{12}}(cX_2^n + U_2) - (cX_2^n + U_2).
\end{align}

The subtractive dither quantization noise $e_{q_i}$ is independent
of both sources $X_1$ and $X_2$ and has the same distribution as
$-U_i$ for $i=1,2$ \cite{zamirmulti}. Since the dithers $U_1$ and
$U_2$ are independent and for a fixed choice of the nested lattice
$e_{q_i}$ is a function of $U_i$ alone, $e_{q_1}$ and $e_{q_2}$ are
independent as well.

Let $e_q = e_{q_1}-e_{q_2}$ be the effective dither quantization
noise. The decoder reconstruction in equation
(\ref{eq:decoderfinal}) can be simplified as
\begin{align}
\hat{Z} &= \left(\frac{\sigma_Z^2-D}{\sigma_Z^2}\right)([Z^n+e_q]
\text{ mod } \Lambda_2) \\ &\stackrel{c.d}{=}
\left(\frac{\sigma_Z^2-D}{\sigma_Z^2}\right) (Z^n+e_q) \label{eq:cdeq} \\
&= Z^n + \left(\left(\frac{\sigma_Z^2-D}{\sigma_Z^2}\right)e_q -
\frac{D}{\sigma_Z^2} Z^n \right) \\ &\triangleq Z^n + N
\label{eq:Ndef}.
\end{align}

The $\stackrel{\text{c.d}}{=}$ in equation (\ref{eq:cdeq}) stands
for equality under the assumption of correct decoding. Decoding
error occurs if equation (\ref{eq:cdeq}) doesn't hold. Let $P_e$ be
the probability of decoding error. Assuming correct decoding, the
distortion achieved by this scheme is the second moment per
dimension\footnote{We refer to this quantity also as the normalized
second moment of the random vector $N$. This should not be confused
with the normalized second moment of a lattice as defined in
equation (\ref{eq:latnorm2mom}).} of the random vector $N$ in
equation (\ref{eq:Ndef}). This can be expressed as
\begin{align}
\frac{\mathbb{E} \parallel N \parallel^2}{n} &=
\left(\frac{\sigma_Z^2-D}{\sigma_Z^2}\right)^2
\frac{\mathbb{E}\parallel e_q \parallel^2}{n} +
\left(\frac{D}{\sigma_Z^2} \right)^2 \frac{\Bbb{E} \parallel Z^n
\parallel^2}{n}
\end{align}
where we have used the independence of $e_{q_1}$ and $e_{q_2}$ to
each other and to the sources $X_1$ and $X_2$ (and therefore to $Z =
X_1 - c X_2$). Since $e_{q_i}$ has the same distribution as $-U_i$,
their expected norm per dimension is just the second moment of the
corresponding lattice $\sigma^2(\Lambda_{1i})$. Thus the effective
distortion achieved by the scheme is
\begin{align}
\nonumber \frac{1}{n}\mathbb{E}\|Z^n-\hat{Z}\|^2 &= \left(
\frac{\sigma_Z^2-D}{\sigma_Z^2} \right)^2 \left(
\frac{D\sigma_Z^2}{\sigma_Z^2-D} \right) + \frac{D^2
\sigma_Z^2}{\sigma_Z^4} \\ &= D \label{eq:Dproof}.
\end{align}
Hence, the proposed scheme achieves the desired distortion provided
correct decoding occurs at equation (\ref{eq:cdeq}). Let us now
prove that equation (\ref{eq:cdeq}) indeed holds with high
probability for an optimal choice of the nested lattice, i.e., there
exists a nested lattice code for which $P_e \rightarrow 0$ as $n
\rightarrow \infty$ where,
\begin{align} \label{eq:pedef}
P_e &= Pr\left( (Z^n+e_q) \text{ mod } \Lambda_2 \neq (Z^n+e_q)
\right).
\end{align}

To this end, let us first compute the normalized second moment of
$(Z^n+e_q)$.
\begin{align}
\frac{\mathbb{E}\parallel Z^n+e_q \parallel^2}{n} &=
\frac{\mathbb{E}\parallel Z^n \parallel^2}{n} + \frac{\mathbb{E}
\parallel -U_1-U_2 \parallel^2}{n} \\ &= \sigma_Z^2 + q_1 + \frac{\sigma_Z^2 D}{\sigma_Z^2-D}-q_1
\\ &= \frac{\sigma_Z^4}{\sigma_Z^2-D} = \sigma^2(\Lambda_2).
\label{eq:channelvariance}
\end{align}

It was shown in \cite{zamirlqn} that as $n \rightarrow \infty$, the
quantization noises $e_{q_i}$ tend to a white Gaussian noise for an
optimal choice of the nested lattice. The following lemma states
that $e_q$ also converges in the same way.
\begin{lemma}
If the two independent subtractive dither quantization noises
$e_{q_i}$ tend to a white Gaussian noise of the same variance as
$e_{q_i}$ in the Kullback-Leibler divergence sense, then $e_q =
e_{q_1} - e_{q_2}$ also tends to a white Gaussian noise of the same
variance as $e_q$ in the divergence sense.
\end{lemma}
\textbf{Proof:} The proof of convergence to Gaussianity of $e_q$ is
detailed in Appendix \ref{sec:eqconvproof}.

We choose $\Lambda_2$ to be an exponentially good channel code in
the sense defined in Section \ref{subsec:lattices} (also see
\cite{zamirmulti}). For such lattices, the probability of decoding
error $P_e$ in equation (\ref{eq:pedef}) goes to $0$ exponentially
fast if $(Z^n+e_q)$ is Gaussian. The analysis in \cite{zamirsnr}
also showed that if $(Z^n+e_q)$ tends to a white Gaussian noise
vector, the effect on $P_e$ of the deviation from Gaussianity is
sub-exponential. Hence, the overall error behavior is asymptotically
the same as the behavior if $(Z^n+e_q)$ were Gaussian, i.e., $P_e
\rightarrow 0$ as $n \rightarrow \infty$. This implies that the
reconstruction error $Z^n-\hat{Z}$ tends in probability to the
random vector $N$ defined in equation (\ref{eq:Ndef}). Since all
random vectors involved have finite normalized second moment, this
convergence in probability implies convergence in second moment as
well. Thus the normalized second moment of the reconstruction error
tends to that of $N$ which is shown to be $D$ in equation
(\ref{eq:Dproof}). Averaged over the random dithers $U_1$ and $U_2$,
we have shown that the appropriate distortion is achieved. Hence
there must exist a pair of deterministic dithers that also achieve
the given distortion. Combining equations (\ref{eq:Rsumproof}) and
(\ref{eq:Dproof}), we have proved the claim of Theorem
\ref{thm:latticerate}.

\textbf{Remark:} Instead of focussing on the entire rate region, if
one is interested in minimizing the sum rate of the encoders, then
it can be checked that the optimal choice of lattice parameters is
$\sigma^2(\Lambda_{11}) = \sigma^2(\Lambda_{12}) = \frac{1}{2}
\frac{D \sigma_Z^2}{\sigma_Z^2-D}$. In this case, we require only
one nested lattice $(\Lambda_1,\Lambda_2)$ with both encoders using
the same nested lattice for encoding.

\subsection{Intuition about the Coding Scheme}
\label{sec:justification} In this section, we outline some arguments
that justify our choice of lattice codes and the scaling constants
described in the previous subsection. Our use of lattice codes is
motivated by the following. Suppose there exists a centralized
encoder that has access to both sources $X_1$ and $X_2$. Clearly,
the optimal encoding strategy then would be to compute $Z=X_1-cX_2$,
and compress it using an encoder, say $f(\cdot)$, that achieves the
optimal rate distortion function of a Gaussian source of variance
$\sigma_Z^2$. Such a centralized coding scheme can be adapted to a
distributed setting if the encoder $f(\cdot)$ \emph{distributes}
over the linear function $X_1-cX_2$. For then, from the decoder's
perspective, there is no distinction between the centralized and
distributed coding scheme since
\begin{align} \label{eq:funcform}
f(X_1-cX_2) &= f(X_1) - f(cX_2).
\end{align}
A lattice code satisfies the functional form mentioned in equation
(\ref{eq:funcform}) and is known to achieve the optimal rate
distortion function for Gaussian sources. Hence it is an ideal
candidate for use as the source encoder.

The parameters of the lattice code as given in equations
(\ref{eq:latsig1}) and (\ref{eq:latsig2}) can be justified as below.
Without loss of generality, let the second source alone be scaled by
an arbitrary constant $\eta$. Let the fine lattices in the signal
path of the two sources have second moments $q_i \triangleq
\sigma^2(\Lambda_{i,1})$ for $i = 1,2$. For the case of optimal
lattices in high enough dimensions, one can think of quantization
using the fine lattices $\Lambda_{i,1}, i=1,2$ as simulating an AWGN
channel of noise variance $q_i$. Such a statement can be made
precise by analysis similar to the one carried out in the previous
subsection. Let $Q_i, i=1,2$ be $\mathcal{N}(0,q_i)$ random
variables that are single-letter asymptotic equivalents of the
subtractive dither quantization noises $e_{q_i}$ encountered in the
previous subsection.

Referring to the equivalent coding scheme represented in Fig.
\ref{fig:eqrep}, we see that it suffices to choose the coarse
lattice $\Lambda_2$ to be a good AWGN channel code of second moment
equal to
\begin{align}
\nonumber \sigma^2(\Lambda_2) &= \text{Var}(X_1+Q_1-(\eta X_2+Q_2)) \\
&= 1+\eta^2-2\eta \rho+q_1+q_2.
\end{align}
Using the distributive property of lattices (equation
(\ref{eq:latdistributive})), this scheme can be converted to the one
represented by Fig. \ref{fig:fullrep}.

The rates achieved by this scheme are given by
\begin{align}
R_i &= \frac{1}{2} \log \frac{1+\eta^2-2\eta \rho + q_1 + q_2}{q_i}
\quad \mbox{for } i =1,2
\end{align}
This region can be optimized over all choices of $\eta$ subject to
an appropriate distortion constraint. It turns out that the scaling
chosen in Section \ref{subsec:mainscheme} is the optimal choice. The
details are described (for the more general $K$ user case) in
Appendix \ref{sec:optchoices}.

\section{Distributed source coding for the $K$ source case} \label{sec:generalizations}

In this section, we consider the case of reconstructing a linear
function of an arbitrary number of sources. In the case of two
sources, the two strategies used in Theorems \ref{thm:latticerate}
and \ref{thm:bergertungrate} were direct reconstruction of the
function $Z$ and estimating the function from noisy versions of the
sources respectively. Henceforth, we shall refer to the coding
scheme used to derive Theorem \ref{thm:latticerate} as lattice
binning  and that used in Theorem \ref{thm:bergertungrate} as random
binning.

In the presence of more than two sources, a host of strategies which
are a combination of these two strategies become available. For
example, in the case of $3$ sources, one possible strategy would be
for all users to use the lattice binning  while another strategy
would be for users 1 and 2 to use lattice binning and user 3 to
employ random binning. The union of the rate-distortion tuples
achieved by all such schemes gives an achievable rate region of the
problem.

When a combination of the two strategies are used among the $K$
sources, the order of decoding at the decoder becomes significant.
The indices which are decoded earlier can be used as side
information for the indices which are to be decoded later. This
raises the question of how to adapt the coding schemes of lattice
binning and random binning to the case when side information is
present at the decoder. For ease of exposition and understanding in
the following section, we first describe a  lattice coding strategy
for the distributed source coding problem involving two sources with
the goal of reconstruction of their linear function at the decoder
and, in addition, the decoder has access to  some side information.
We then use this to formally describe an achievable rate region for
the problem of reconstructing $Z = \sum_{i=1}^{K} c_i X_i$.

\subsection{Lattice coding in presence of decoder side information}
\label{subsec:latcodingsideinfo} In this section, we consider the
problem of distributed encoding of correlated sources using lattices
in the presence of side information at the decoder. As we will see,
this can be used as a building block in reconstructing a linear
function of multiple sources.

Assume that we have correlated Gaussian sources $X_1$ and $X_2$ and
the decoder is interested in reconstructing a linear function $Z
\triangleq \sum_{i=1}^{2} c_i X_i$. Suppose the decoder also has
available to it side information $Y$ that is correlated with the
sources $X_1,X_2$. $Y$ and $X_1,X_2$ are jointly Gaussian. Each
source $X_i$ is observed by an encoder which maps its outcomes to a
finite set. The indices produced by the encoders are transmitted to
a joint decoder using two rate-constrained noiseless channels. The
goal is to find the optimal rate-distortion region which is the set
of all achievable tuples $(R_1,R_2,D)$.

In this subsection we provide an inner bound to the optimal
rate-distortion region for this problem using a lattice-based
``correlated'' binning strategy.   We use the notation $\hat{Z}_{Y}$
to denote the minimum mean-squared error (MMSE) estimate of $Z$
given $Y$, namely $\mathbb{E}(Z \mid Y)$. The innovations random
variable $Z - \hat{Z}_{Y}$ is denoted by $\eta_{Z \mid
  Y}$.

The lattice coding strategy in the presence of side information can
be inferred by considering what the strategy would be in the
presence of a central encoder that has access to all the sources
$X_1,X_2$ and the side information $Y$. In that case, the central
encoder would first compute $Z = \sum_{i=1}^{2} c_i X_i$ and then
quantize and transmit only the innovations random variable $\eta_{Z
\mid Y}$. This can be accomplished with subtractive dither lattice
quantization using a nested lattice $\Lambda_2 \subset \Lambda_1$ of
parameter
\begin{align}
\sigma^2(\Lambda_1) &= \frac{D \sigma_{\eta}^2}{\sigma_{\eta}^2 - D}
\\ \sigma^2(\Lambda_2) &= \frac{ \sigma_{\eta}^4} { \sigma_{\eta}^2
- D}
\end{align}
where $\sigma^2_{\eta}$ is the variance of the innovations random
variable $\eta_{Z\mid Y}$ and $D$ is the desired distortion in the
reconstruction of $Z$. The rate incurred in this system is given by
$\frac{1}{2} \log (\sigma_{\eta}^2/D)$. The decoder would use this
quantized innovations with the side information to obtain a
reconstruction that is within a distortion of $D$ of $Z$.

The two assumptions in the setup above that deviate from our
distributed coding problem are that all sources are available to a
central encoder and that side information is available at the
encoder. The first assumption can be gotten rid of by employing the
distributive property (equation (\ref{eq:latdistributive})) of
lattice codes. The second assumption can be eliminated by using the
linear nature of the forward test channel for the case of Gaussian
quantization. This linear nature enables one to move the side
information present at the encoder to the decoder thus obviating its
necessity at the encoder. Thus, we can convert the above centralized
coding strategy to our distributed setting to yield the following
encoding scheme.

The source encoders are described by the equations
\begin{align}
S_i &= (Q_{\Lambda_{1i}}(c_i X_i^n + U_i)) \text{ mod } \Lambda_2
\quad \text{for } i = 1,2, \label{sideinfo_encoding}
\end{align}
where $U_i$s are independent random dithers uniformly distributed
over the fundamental Voronoi region $\mathcal{V}_{0,1i}$ of the fine
lattices $\Lambda_{1i}$s. As in Section \ref{sec:defandresult}, we
require $\Lambda_2 \subset \Lambda_{1i}, \, i = 1,2$, the fine
lattices $\Lambda_{1i}$ to be good source codes and the coarse
lattice $\Lambda_2$ to be a good channel code. The second moments of
the nested lattices are given by
\begin{align}
\sigma^2(\Lambda_{11}) &= q_1 \label{eq:sideinfoenc1}\\
\sigma^2(\Lambda_{12}) &= \frac{D \sigma_{\eta}^2}{\sigma_{\eta}^2 -
D} - q_1 \label{eq:sideinfoenc2}
\\ \sigma^2(\Lambda_2) &= \frac{ \sigma_{\eta}^4} { \sigma_{\eta}^2 -
D} \label{eq:sideinfoenc3}
\end{align}
where $q_1$ is chosen such that $0 < q_1 <
\frac{D\sigma_{\eta}^2}{\sigma_{\eta}^2 - D}$. This gives a
quantization rate of
\begin{align}
R_1 &= \frac{1}{2} \log
\frac{\sigma_{\eta}^4}{q_1(\sigma_{\eta}^2-D)}
\label{eq:sideinforate1} \\
R_2 &= \frac{1}{2} \log
\frac{\sigma_{\eta}^4}{D\sigma_{\eta}^2-q_1(\sigma_{\eta}^2-D)}
\label{eq:sideinforate2}
\end{align}
Clearly, for a fixed choice of $q_1$ all rates beyond that given
above can be achieved. Eliminating $q_1$ between the two rates now
gives us an expression of the overall achievable region as
\begin{align} \label{eq:sideinforateregion}
2^{-2R_1} + 2^{-2R_2} &\leq \left(\frac{\sigma_{\eta}^2}{D}
\right)^{-1}
\end{align}

The decoder is given by the equation
\begin{align} \label{eq:sideinfodecdesc}
\hat{Z} &= \left(1 - \frac{D}{\sigma_{\eta}^2}\right) \left(
\left[\sum_{i=1}^{2} (S_i-U_i)- \hat{Z}_{Y}^n \right] \text{ mod }
\Lambda_2 \right) + \hat{Z}_{Y}^n
\end{align}

The encoding operation given by equation (\ref{sideinfo_encoding})
is similar to that used in Section \ref{subsec:mainscheme}. By
mimicking the analysis of Section \ref{subsec:mainscheme}, we can
show that the first part of the decoder operation, given by
$([\sum_{i=1}^{2}(S_i-U_i) -\hat{Z}_{Y}^n] \text{ mod } \Lambda_2)$
in equation (\ref{eq:sideinfodecdesc}), produces with high
probability $\eta_{Z\mid Y}^n+N$ where $N$ approaches a white
Gaussian noise vector with each element having variance
$\sigma^2(\Lambda_{11}) + \sigma^2(\Lambda_{12}) =
\frac{D\sigma_{\eta}^2}{\sigma_{\eta}^2-D}$. The decoder then
obtains an estimate of the function $Z$ based on $\eta_{Z\mid Y}+N$
and the side information $Y$. It can be checked that equation
(\ref{eq:sideinfodecdesc}) describes such an estimate and that this
estimate indeed achieves the desired distortion $D$. Thus, we have
an achievable rate-distortion tuple given by equation
(\ref{eq:sideinforateregion}) for reconstructing a linear function
in the presence of any side information. The rationale for choosing
the lattice parameters and scaling constants is very similar to that
given in Section \ref{sec:justification}.

\subsection{Reconstructing a linear function of $K$ sources}
\label{subsec:Kcase} Previously, we considered the problem of
reconstructing a linear function of two sources. In this section, we
generalize the problem to an arbitrary number of sources. Let the
sources be given by $X_1,X_2,\ldots, X_K$ which are jointly
Gaussian. The encoder of $X_i$ maps its outcome to a finite set. The
output of the encoder is transmitted over a noiseless but
rate-constrained channel to a joint decoder. The rate of channel $i$
is given by $R_i$. The decoder wishes to reconstruct a linear
function given by $Z=\sum_{i=1}^K c_iX_i$ with squared error
fidelity criterion. The performance limit $\mathcal{RD}$ is given by
the set of all rate-distortion tuples $(R_1,R_2,\ldots,R_K,D)$ that
are achievable in the sense defined in Section
\ref{sec:defandresult}. In this section we provide an inner bound
based on ``correlated'' lattice-structured binning.

As indicated earlier, there are several possible coding schemes based
on each user's choice of coding strategy and also the choice of
order of decoding. Before, we describe these coding schemes, we
introduce some relevant notation.

For any set $A \subset \{1,\dots,K\}$, let $X_A$ denote those
sources whose indices are in $A$, i.e., $X_A \triangleq \{X_i \, :
\, i \in A \}$. Let $Z_A$ be defined as $\sum_{i \in A} c_i X_i$.
Let $\Theta$ be a partition of $\{1,\dots,K\}$ with $\theta =
|\Theta|$. Let $\pi_{\Theta} : \Theta \rightarrow
\{1,\dots,\theta\}$ be a permutation. One can think of
$\pi_{\Theta}$ as ordering the elements of $\Theta$. Each set of
sources $X_A, A \in \Theta$ are decoded simultaneously at the
decoder with the objective of reconstructing $Z_A$. The order of
decoding is given by $\pi_{\Theta}(A)$ with the lower ranked sets of
sources decoded earlier. Let $\mathcal{Q} = (q_1,\ldots,q_K) \in
\Bbb{R}_{+}^{K}$ be a tuple of positive reals. Let $\Bbb{E}(\cdot)$
denote the expectation operator.

For any partition $\Theta$ and ordering $\pi_{\Theta}$, let us
define recursively a positive-valued function $\sigma^2_{\Theta}:
\Theta
 \rightarrow \Bbb{R}^+$ as follows:
\beq \label{eq:sigAdef} \sigma^2_{\Theta}(A)=\Bbb{E} \left[ (Z_A -
f_{A}(S_A))^2 \right], \eeq where \beq f_A(S_A) = \Bbb{E}(Z_A|S_A)
\label{eq:fAdef} \eeq \beq S_A = \{Z_{B}+Q_B: B \in \Theta,
\pi_{\Theta} (B) < \pi_{\Theta} (A) \} \label{eq:sAdef} \eeq and
$\{Q_A: A \in \Theta\}$ is a collection of $|\Theta|$ independent
zero-mean Gaussian random variables with variances given by $q_A =
\text{Var}(Q_A) \triangleq \sum_{i \in A} q_i$, and this collection
is independent of the sources. Let \beq f(\{Z_A+Q_A:A \in \Theta\})
\triangleq \Bbb{E}\left(Z| \{Z_A+Q_A:A \in \Theta\} \right).
\label{eq:Fdef} \eeq

\begin{thm} \label{thm:Ksrcthm}
For a given tuple of sources $X_1,\ldots,X_K$ and tuple of real
numbers $(c_1,c_2,\ldots,c_K)$, we have $\mathcal{RD}_{in}^* \subset
\mathcal{RD}$, where \beq \nonumber \mathcal{RD}_{in} =
\bigcup_{\Theta,\pi_{\Theta},\mathcal{Q}} \left\{(R_1,\ldots,R_K,D):
R_i \geq \frac{1}{2} \log \frac{ \sigma^2_{\Theta}(A) + q_A}{q_i}
\mbox{ for } i \in A \label{eq:Ksrcsumrate} \right. \eeq \beq \left.
D \geq \Bbb{E}\left[(Z- f \left( \{Z_A+Q_A:A \in \Theta\}
\right)\right)^2] \ \ \ \right\}, \label{eq:Ksrcdistortion} \eeq and
$^*$ denotes convex closure.
\end{thm}

\textbf{Proof:} We give a description of a lattice-based coding
scheme that achieves the inner bound. Fix $\Theta$, $\pi_{\Theta}$
and $\mathcal{Q}$. For each $A \in \Theta$, construct a family of
good nested lattices $\Lambda_{1i}^A$ and $\Lambda_2^A$ such that
$\Lambda_2^A \subset \Lambda_{1i}^A$ for $i \in A$. Existence of
such good nested lattices has been shown in Appendix
\ref{sec:nestedlatticeproof}. The second moment of the fine lattice
$\Lambda_{1i}^A$ is chosen to be $q_i$. The second moment of the
coarse lattice is chosen based on the amount of side information
available to the decoder at the time of decoding the set of sources
$X_A$ which in turn depends on $\pi_{\Theta}(A)$. The function
$\sigma^2_{\Theta}$ governs this choice. More precisely, for $i \in
A$ and $A \in \Theta$, the second moments of the lattices are given
by
\begin{align}
\sigma^2(\Lambda_{1i}^A) &= q_i \label{eq:Ksrcenc1} \\
\sigma^2(\Lambda_2^A) &= \sigma^2_{\Theta}(A)+ q_A
 \label{eq:Ksrcenc2}
\end{align}

\emph{Encoder:} For each $A \in \Theta$, the source $X_i$, $i \in A$
is encoded using the nested lattice $\Lambda_2^A \subset
\Lambda_{1i}^A$. The encoders can be described by the equations
\begin{align} \label{eq:Ksrcenceq}
T_i &= (Q_{\Lambda_{1i}^A}(c_i X_i^n + U_i)) \text{ mod }
\Lambda_2^A \quad \text{for } i \in A
\end{align}
where $U_i$ are independent random dithers uniformly distributed
over the fundamental Voronoi region $\mathcal{V}_{0,{1i}}^A$ of the
fine lattice $\Lambda_{1i}^A$. This would give an encoding rate of
\begin{align} \label{eq:Ksrcrate}
R_i &= \frac{1}{2} \log \frac{ \sigma^2_{\Theta}(A)+q_A}{q_i} \quad
\text{for } i \in A
\end{align}

\emph{Decoder:} For $A \in \Theta$, in order to decode $Z_A$, the
decoder has access to some side information and its operation can be
recursively described similar to equations (\ref{eq:decdesc}) and
(\ref{eq:sideinfodecdesc}) as
\begin{align} \label{eq:Ksrcdeceq}
\hat{Z}_A &= \left(
\left[\sum_{i \in A} (T_i - U_i)  - f_A^n(\hat{S}_A)  \right] \text{ mod } \Lambda_2^A
\right) + f_A^n(\hat{S}_A)
\end{align}
where \beq \hat{S}_A=\{\hat{Z}_{B}: B \in \Theta, \pi_{\Theta} (B) <
\pi_{\Theta} (A) \}. \label{eq:hatsAdef} \eeq After decoding
$\hat{Z}_A$ for all $A \in \Theta$, the decoder obtains the
reconstruction as a linear function of $\{\hat{Z}_A: A \in \Theta\}$
as \beq \hat{Z}= f^n(\{\hat{Z}_A:A \in \Theta\}). \eeq

We now show that the above system achieves the inner bound given in
the theorem. From equation (\ref{eq:Ksrcrate}), it is clear that
this scheme achieves the rate tuple claimed in Theorem
\ref{thm:Ksrcthm}. It remains to prove that the claimed distortion
is achieved. The crucial observation is that while $S_A$ in equation
(\ref{eq:sAdef}) denotes the side information available to decode
$Z_A$ in test channels, $\hat{S}_A$ in equation (\ref{eq:hatsAdef})
denotes the side information available to decode $\hat{Z}_A$ in the
actual coding system. If we were to assume $\hat{S}_A$ to be
Gaussian, then by definition of the functions $f_A(\cdot)$ (equation
(\ref{eq:fAdef})) and $f(\cdot)$ (equation (\ref{eq:Fdef})), it is
easy to see that the distortion given in Theorem \ref{thm:Ksrcthm}
is achieved. However such an assumption isn't true for $\hat{S}_A$
for any finite lattice dimension $n$.

Fortunately, loosely speaking, we can show that even though the
assumption of Gaussianity of $\hat{Z}_A$ isn't strictly true, it
becomes increasingly valid as the lattice dimension $n \rightarrow
\infty$. By analysis similar to that in Section
\ref{subsec:mainscheme}, we can show that the subtractive dither
quantization noises tend to a white Gaussian of the same variance
(in the K-L divergence sense). This implies that as the lattice
dimension $n \rightarrow \infty$, for an optimal choice of nested
lattices, $\hat{Z}_A$ tends to $Z_A^n + Q_A^n$ and hence $\hat{S}_A$
tends to $S_A^n$ (in the K-L divergence sense). By virtue of the
``goodness'' of the nested lattices, this then implies that the
probability of incorrect decoding goes to $0$ exponentially in the
lattice dimension. Thus the reconstruction error $(Z^n-\hat{Z})$
tends in probability (and hence in normalized second moment) to $N$
where $N$ approaches a Gaussian random vector with each component
having variance $D$. Thus, the proposed lattice scheme indeed
achieves the claimed rate-distortion tuples and Theorem
\ref{thm:Ksrcthm} is proved.

To show this formally using induction, we need some more notation.
For each $A \in \Theta$ and for each $i \in A$, let \beq
e_i=Q_{\Lambda_{1i}^A}(c_iX_i^n+U_i) -c_iX_i^n-U_i, \eeq and \beq
e_A \triangleq \sum_{i \in A} e_i. \eeq For each $A \in \Theta$, let
the linear function $f_A(\cdot)$ be given by \beq f_A(S_A)=\sum_{B:
\pi_{\Theta}(B) < \pi_{\Theta}(A)} \alpha_A(B) (Z_B+Q_B). \eeq By
noting that $e_i$ are independent for $i \in \{1,2,\ldots,K\}$, we
note that for all $A \in \Theta$, \beq \frac{1}{n} \Bbb{E} \| e_A
\|^2= q_A. \eeq Let $E \in \Theta$ be such that $\pi_{\Theta}(E)=1$.
Thus $\hat{S}_E=\phi$. Hence using the distributive property, and
noting the normalized second moments of $e_i$ for $i \in E$, we have
with high probability (i.e., under correct decoding) \beq \hat{Z}_E
= Z_E^n+e_E. \eeq For any $1\leq j<K$, we assume correct decoding
with high probability at the $j$th stage and show  correct decoding
with high probability at the $(j+1)$th stage. Let $C \in \Theta$ be
such that $\pi_{\Theta}(C)=j+1$. Under the above assumption, we
have, with high probability, for all $B \in \Theta$ with
$\pi_{\Theta}\leq j$ \beq \hat{Z}_B=Z_B^n+e_B. \eeq Using this we
have
\begin{eqnarray}
\hat{Z}_C &=& \left( Z_C^n+e_C- \sum_{B: \pi_{\Theta}(B)\leq j } \alpha_C(B)
\hat{Z}_B \right) \ \mbox{ mod } \   \Lambda_2^C+ \sum_{B:
  \pi_{\Theta}(B) \leq j} \alpha_C(B)
\hat{Z}_B \\
&\stackrel{c.d}{=}& \left( Z_C^n +e_C- \sum_{B: \pi_{\Theta}(B) \leq j} \alpha_C(B)
\hat{Z}_B \right) + \sum_{B: \pi_{\Theta}(B) \leq j } \alpha_C(B)
\hat{Z}_B \\
&=&Z_C^n+e_C,
\end{eqnarray}
where the second equality holds with high probability (correct
decoding) because of the following reasons.  (a) The normalized
second moment of the term inside the mod operation satisfies the
following equalities: \beq \frac{1}{n} \Bbb{E} \left\| Z_C^n +e_C-
\sum_{B: \pi_{\Theta}(B) \leq j} \alpha_C(B) \hat{Z}_B \right\|^2 =
\hspace{3in} \eeq
\begin{eqnarray}
\hspace{0.5in} &=& \frac{1}{n} \Bbb{E} \left\|Z_C^n - \sum_{B:
\pi_{\Theta}(B)
  \leq j} \alpha_C(B)  Z_B^n  \right\|^2+q_C+
\sum_{B: \pi_{\Theta}(B)  \leq j} \alpha_C^2(B) q_B \\
&=& q_C+ \Bbb{E} \left( Z_C- \sum_{B: \pi_{\Theta}(B) \leq j}
\alpha_C(B)
(Z_B+Q_B) \right)^2  \\
&=& \sigma^2_{\Theta}(C)+q_C \\ &=& \sigma^2(\Lambda_2^C).
\end{eqnarray}
(b) Using the arguments of Section \ref{subsec:mainscheme} (see
Appendix \ref{sec:eqconvproof}), \beq \lim_{n \rightarrow \infty} h
\left( Z_C^n+e_C- \sum_{B: \pi_{\Theta}(B) \leq j} \alpha_C(B)
\hat{Z}_B \right) = \frac{n}{2} \log 2 \pi e \sigma^2(\Lambda_2^C).
\eeq where $h(\cdot)$ denotes differential entropy. Hence we have
for all $A \in \Theta$, with high probability, \beq
\hat{Z}_A=Z_A^n+e_A. \eeq Now regarding the final estimation, an
argument similar to the above can be given that shows that a
distortion given in the theorem is achieved asymptotically. The
rationale for the specific choice of scaling constants is explained
in detail in Appendix \ref{sec:optchoices}.

~\\
\textbf{Remark:} An important point worth noting before proceeding
further is that the nesting relations we need the lattices to
satisfy is $\Lambda_{2}^A \subset \Lambda_{1i}^A$ for $i \in A$.
But, for $A,B \in \Theta$, we don't need the lattice families
$(\Lambda_{1i}^A,\Lambda_2^A)$ and $(\Lambda_{1j}^B,\Lambda_2^B)$ to
be related in any way for $A \neq B$. Also, just as in the two user
case, if we are interested only in minimizing the sum rate of this
encoding scheme, then for all encoders in a given set $A \in
\Theta$, the second moment of their respective fine lattices are
equal. This means that all encoders in a given set $A \in \Theta$
can use the same nested lattice $\Lambda_2^A \subset \Lambda_1^A$
for encoding.

\subsection{An illustration of Theorem \ref{thm:Ksrcthm}}

For clarity, an illustration of the coding scheme of Theorem
\ref{thm:Ksrcthm} for the case of $6$ users and specific choices of
$\Theta$ and $\pi_{\Theta}$ is described below. Let us choose
$\Theta = \{ \{1,2,3\},\{4,5\},\{6\} \}$. Let $\pi_{\Theta}$ be the
identity permutation so that $\pi_{\Theta}(\{1,2,3\}) = 1,
\pi_{\Theta}(\{4,5\}) = 2, \pi_{\Theta}(\{6\}) = 3$. This means that
the decoder decodes $Z_{\{1,2,3\}} = \sum_{i=1}^{3} c_i X_i$ first
which is then used as side information for decoding $Z_{\{4,5\}}$
and so on. Let us also fix $\mathcal{Q} = \{q_1,\ldots,q_6 \}$ where
$q_i$ are all positive. We use $A,B,C$ to denote the sets
$\{1,2,3\}, \{4,5\}$ and $\{6\}$ respectively.

\begin{figure*}[htp]
\centering
\includegraphics[width = \textwidth]{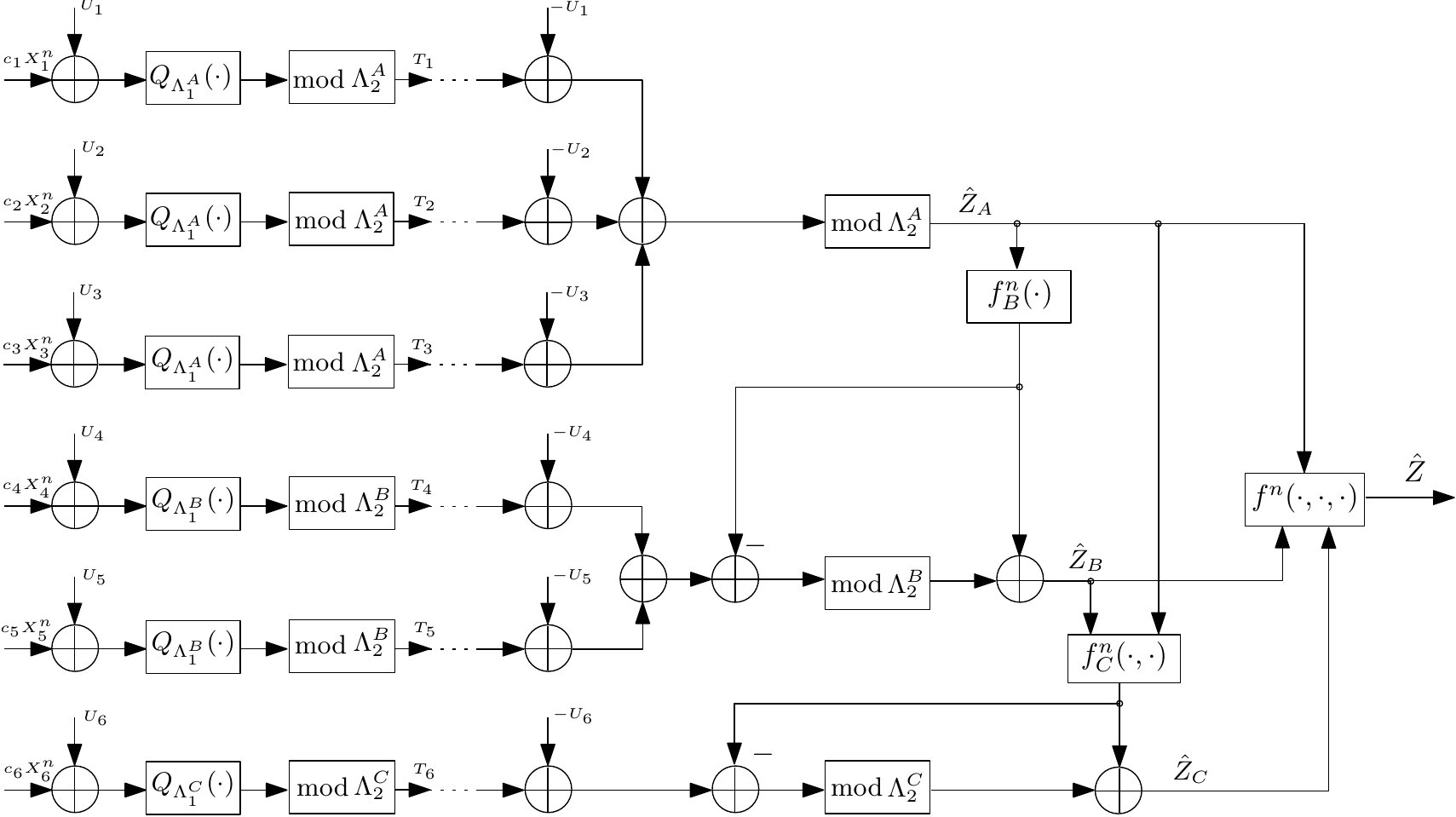}
\caption{Illustration of the coding scheme of Theorem
\ref{thm:Ksrcthm}} \label{fig:Ksrcfig}
\end{figure*}

The fine lattice of the encoder of source $X_i$ has second moment
$q_i$ as given in equation (\ref{eq:Ksrcenc1}). Encoders for the
sources $X_1,X_2,X_3$ use nested lattices where the second moment of
the coarse lattices are given by equation (\ref{eq:Ksrcenc2}). The
decoder decodes $\hat{Z}_A$ according to equation
(\ref{eq:Ksrcdeceq}). To decode $\hat{Z}_A$, the decoder does not
have access to any side information. Encoders for $X_4,X_5$ use
nested lattices whose parameters depend on the function
$\sigma_{\Theta}^2(B)$ which in turn is determined by the fact that
$\hat{Z}_A$ has been decoded earlier. The decoder then decodes
$\hat{Z}_B$ from $T_4,T_5$ and the functional value $f_B^n(\cdot)$
of the side information $\hat{S}_B = \hat{Z}_A$. Similarly, to
decode $\hat{Z}_C$, the decoder has side information $\hat{S}_C = \{
\hat{Z}_A, \hat{Z}_B \}$ along with the index $T_6$. After having
decoded $\hat{Z}_A,\hat{Z}_B,\hat{Z}_C$, the decoder uses the
function $f^n(\cdot)$ of equation (\ref{eq:Fdef}) to estimate $Z$.
This is illustrated in Fig. \ref{fig:Ksrcfig}.

\subsection{A Few Special Cases}
In this section, we consider a few special cases of the general
coding problem treated above. In particular, we examine the rate
distortion region derived above for specific choices of the
partition $\Theta$. First, we demonstrate that we can recover the
two user rate region of Theorems \ref{thm:latticerate} and
\ref{thm:bergertungrate} from the more general $K$-user rate region
described above. Then, we illustrate a scheme  for the case where
the decoder estimates the function directly, i.e., $\Theta=\{
\{1,2,\ldots,K\}\}$.

\subsubsection{Berger Tung coding for the two user case}
In this section, we rederive the result of Theorem
\ref{thm:bergertungrate} using the more general framework of Theorem
\ref{thm:Ksrcthm}. Let the function to be reconstructed be $Z = X_1
- cX_2$ as in Section \ref{sec:defandresult}. Individual
reconstruction of the sources corresponds to the partition $\Theta =
\{ \{1\},\{2\} \}$. There are two possible choices of $\pi_{\Theta}$
corresponding to which source is decoded first. Let us choose
$\pi_{\Theta}$ to be the identity permutation.  Thus $Z_{\{1\}} =
X_1$ is decoded first and used as side information to decode $Z_{
\{2\} } = -cX_2$.

Let $\mathcal{Q} = (q_1,q_2)$ where $q_i$ are positive for $i=1,2$.
For ease of notation, we drop the set notation in the subscripts
below. In what follows, $S_1$ is taken to mean $S_{\{1\}}$ and so
on. Equations (\ref{eq:sigAdef}) to (\ref{eq:sAdef}) simplify in
this case to \beq S_1 = \phi \eeq \beq f_1(S_1) = \Bbb{E}(Z_1) = 0
\eeq \beq \sigma^2_{\Theta}(\{1\}) = \Bbb{E}(Z_1^2) = 1 \eeq \beq
S_2 = \{ X_1 + Q_1 \} \eeq \beq f_2(S_2) = \Bbb{E} (Z_2 \mid S_2) =
\Bbb{E}(-cX_2 \mid X_1 + Q_1) = \frac{-\rho c}{1+q_1} S_2 \eeq \beq
\sigma^2_{\Theta}(\{2\}) = \Bbb{E}\left(Z_2 + \frac{\rho c}{1+q_1}
S_2 \right)^2 = c^2 + q_2 - \frac{\rho^2 c^2}{1+q_1}. \eeq

It follows from estimation theory that the function
$f(Z_1+Q_1,Z_2+Q_2) = a(Z_1+Q_1) + b(Z_2+Q_2)$ where the constants
$a,b$ are given by
\begin{align}
[ \begin{array}{cc} a & b \end{array} ] &= \left[
\begin{array} {cc} \frac{\alpha c^2 + q_2(1-\rho
c)}{(1+q_1)(c^2+q_2) - \rho^2 c^2} & \frac{c(\alpha
c+q_1(c-\rho))}{(1+q_1)(c^2+q_2)-\rho^2 c^2}
\end{array} \right]
\end{align} where $\alpha \triangleq 1-\rho^2$.

As stated in Theorem \ref{thm:Ksrcthm}, $q_i$ have to satisfy the
distortion constraint of equation (\ref{eq:Ksrcdistortion}) which in
this case simplifies to
\begin{align} \label{eq:KsrcBTdist}
D &\geq \frac{q_1 c^2 \alpha + q_2 c^2 \alpha + q_1 q_2
\sigma_Z^2}{(1+q_1)(c^2+q_2) - \rho^2 c^2}
\end{align}
The parameters of the nested lattices are given by equations
(\ref{eq:Ksrcenc1}) and (\ref{eq:Ksrcenc2}) to be
\begin{align}
\sigma^2(\Lambda_1^{\{1\}}) &= q_1 \\
\sigma^2(\Lambda_2^{\{1\}}) &= 1+q_1 \\
\sigma^2(\Lambda_1^{\{2\}}) &= q_2 \\
\sigma^2(\Lambda_2^{\{2\}}) &= c^2 + q_2 - \frac{\rho^2 c^2}{1+q_1}.
\end{align}
This gives the following rates.
\begin{align}
R_1 &= \frac{1}{2} \log \frac{1+q_1}{q_1} \\
R_2 &= \frac{1}{2} \log \frac{(c^2+q_2)(1+q_1) - \rho^2 c^2}{q_2
(1+q_1)}
\end{align}
where $\mathcal{Q} = (q_1,q_2)$ is subject to the distortion
constraint of equation (\ref{eq:KsrcBTdist}). It can be checked that
these equations parameterize one of the corner points of the rate
region of Theorem \ref{thm:bergertungrate}. Reversing the roles of
the two sources (equivalently, choosing $\pi_{\Theta}(\{1\}) = 2,
\pi_{\Theta}(\{2\}) = 1$), we can achieve the other end point of the
rate region. Time sharing between these two points achieves the
entire rate region of Theorem \ref{thm:bergertungrate}.

Note that the inner bound of Theorem \ref{thm:bergertungrate} is
derived using the Berger-Tung inner bound \cite{berger77,bergertung}
which employs random quantization followed by random binning. Here,
we have rederived this result using lattice quantization followed by
lattice-structured binning.

\subsubsection{Lattice coding for the K user case}
\label{subsec:latcodingKuser}

In this section, we derive an achievable rate region for the $K$
user case when all the users encode in such a way that the decoder
estimates the function directly without reconstructing any
intermediate variables. This corresponds to the case where $\Theta =
\{ \{1,\dots,K \} \}$. $\pi_{\Theta}$ is trivial in this case. Let
$\mathcal{Q} = \{q_1,\ldots,q_K\} \in \Bbb{R}_{+}^K$. Let $A$ denote
the set $\{1,\dots,K\}$. Then $q_A = \sum_{i=1}^K q_i$

Equations (\ref{eq:sigAdef}) to (\ref{eq:sAdef}) simplify in this
case to \beq S_A = \phi \eeq \beq f_A(S_A) = \Bbb{E}(Z) = 0 \eeq
\beq \sigma^2_{\Theta}(A) = \Bbb{E}(Z^2) = \sigma_Z^2. \eeq The
function $f(\cdot)$ of equation (\ref{eq:Fdef}) is given by
\begin{align}
\nonumber f({Z+Q}) &= \Bbb{E}(Z \mid Z+Q) \\ &=
\frac{\sigma_Z^2}{\sigma_Z^2+q_A}(Z+Q)
\end{align}
and thus distortion constraint of equation (\ref{eq:Ksrcdistortion})
fixes the value of $q_A$ to be $\frac{\sigma_Z^2 D}{\sigma_Z^2-D}$.

The encoders use the nested lattices $(\Lambda_{1i},\Lambda_2), i =
1,\ldots,K$ for encoding. The parameters of the nested lattices are
given by
\begin{align}
\sigma^2(\Lambda_{1i}) &= q_i \\
\sigma^2(\Lambda_2) &= \sigma_Z^2 + q_A =
\frac{\sigma_Z^4}{\sigma_Z^2-D}
\end{align}
This gives an encoding rate of
\begin{align} \label{eq:Kuserlatencrate}
R_i &= \frac{1}{2} \log \frac{\sigma_Z^4}{q_i(\sigma_Z^2-D)}
\end{align}
This corresponds to the rate region
\begin{align}
\sum_{i=1}^K 2^{-2R_i} &\leq \left( \frac{\sigma_Z^2}{D}
\right)^{-1}
\end{align}
For $K=2$, this recovers the rate region of Theorem
\ref{thm:latticerate}.

\section{Comparison of the Rate Regions} \label{sec:comparison}
In this section, we compare the rate regions of the lattice based
coding scheme given in Theorem \ref{thm:latticerate} and the
Berger-Tung based coding scheme given in Theorem
\ref{thm:bergertungrate} for the case of two users. The function
under consideration is $Z = X_1 - cX_2$. We would like to emphasize
that we have assumed that the sources have unit variance and that
$\rho > 0$. To demonstrate the performance of the lattice binning
scheme, we choose the sum rate of the two encoders as the
performance metric.

\begin{figure}[htp]
\centering
\includegraphics[width = 0.6\textwidth]{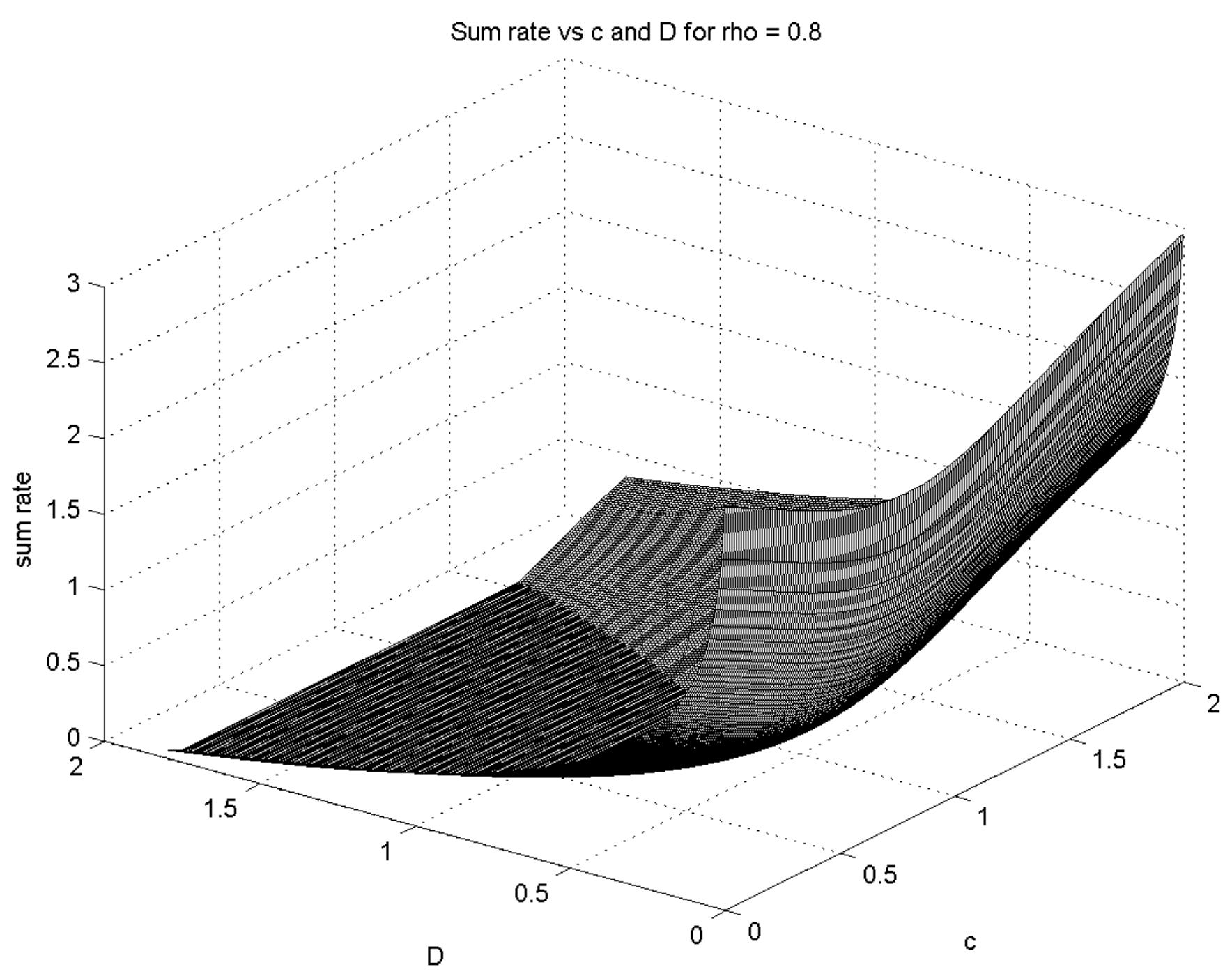}
\caption{Lattice based scheme's sum-rate vs $c$ and distortion $D$
for $\rho = 0.8$} \label {fig:3dplot08}
\end{figure}

\begin{figure}[htp]
\centering
\includegraphics[width = 0.6\textwidth]{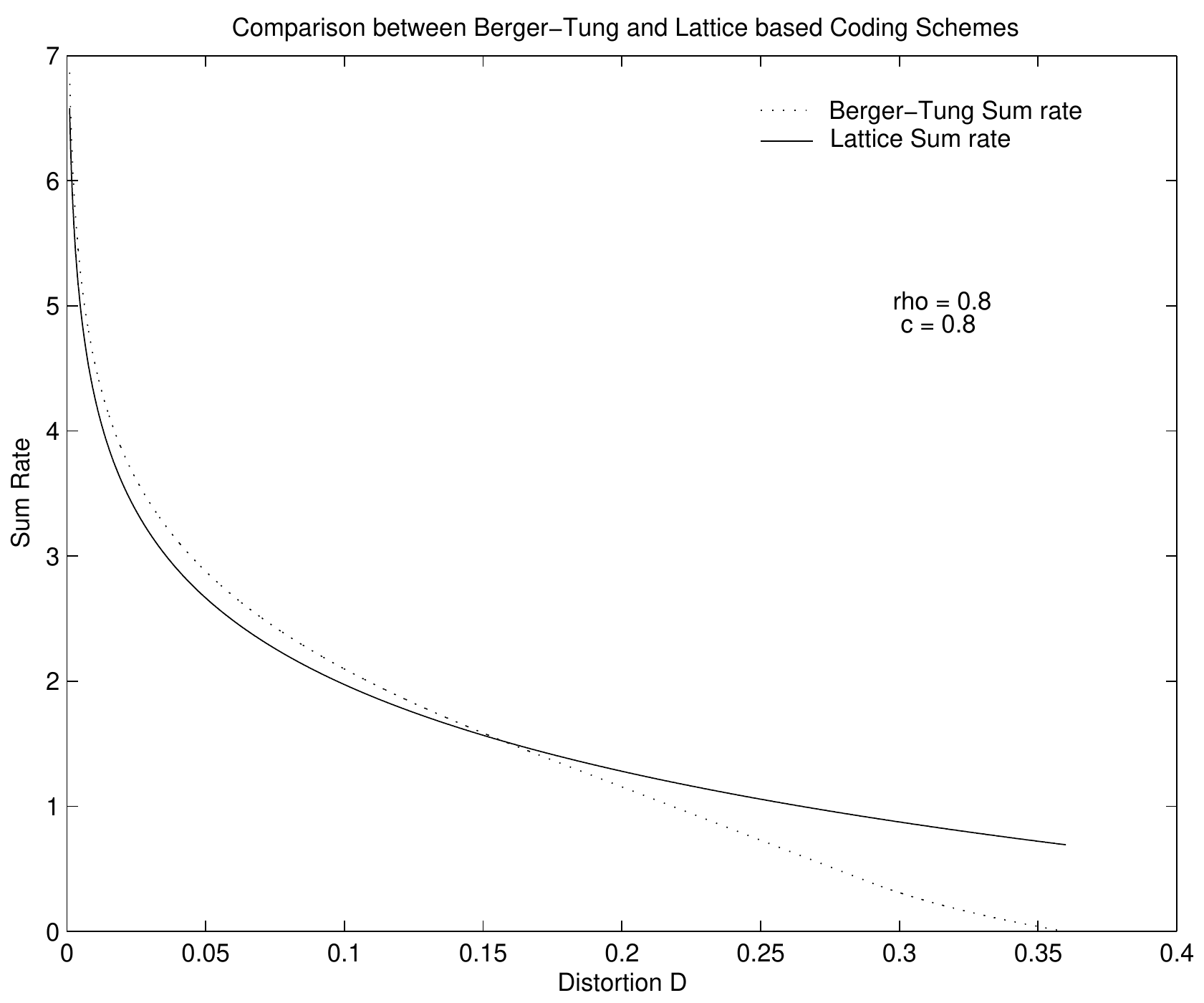}
\caption{Comparison of the sum-rates of the 2 schemes}
\label{fig:rateplot1}
\end{figure}

Fig. \ref{fig:3dplot08} shows the sum rate of the lattice based
scheme for different values of $c$ and distortion $D$. In Fig.
\ref{fig:rateplot1}, we compare the sum-rates of the two schemes for
$\rho=0.8$ and $c=0.8$. Fig. \ref{fig:rateplot1} shows that for
small distortion values, the lattice scheme achieves a smaller sum
rate than the Berger-Tung based scheme. This shows that the rate
region of Theorem \ref{thm:latticerate} contains points outside that
of the rate region of Theorem \ref{thm:bergertungrate}. The opposite
is also true since for $D = \sigma_Z^2$, the region in Theorem
\ref{thm:bergertungrate} contains the rate point $(0,0)$ while the
one in Theorem \ref{thm:latticerate} does not.

\begin{figure}[htp]
\centering
\includegraphics[width = 0.6\textwidth]{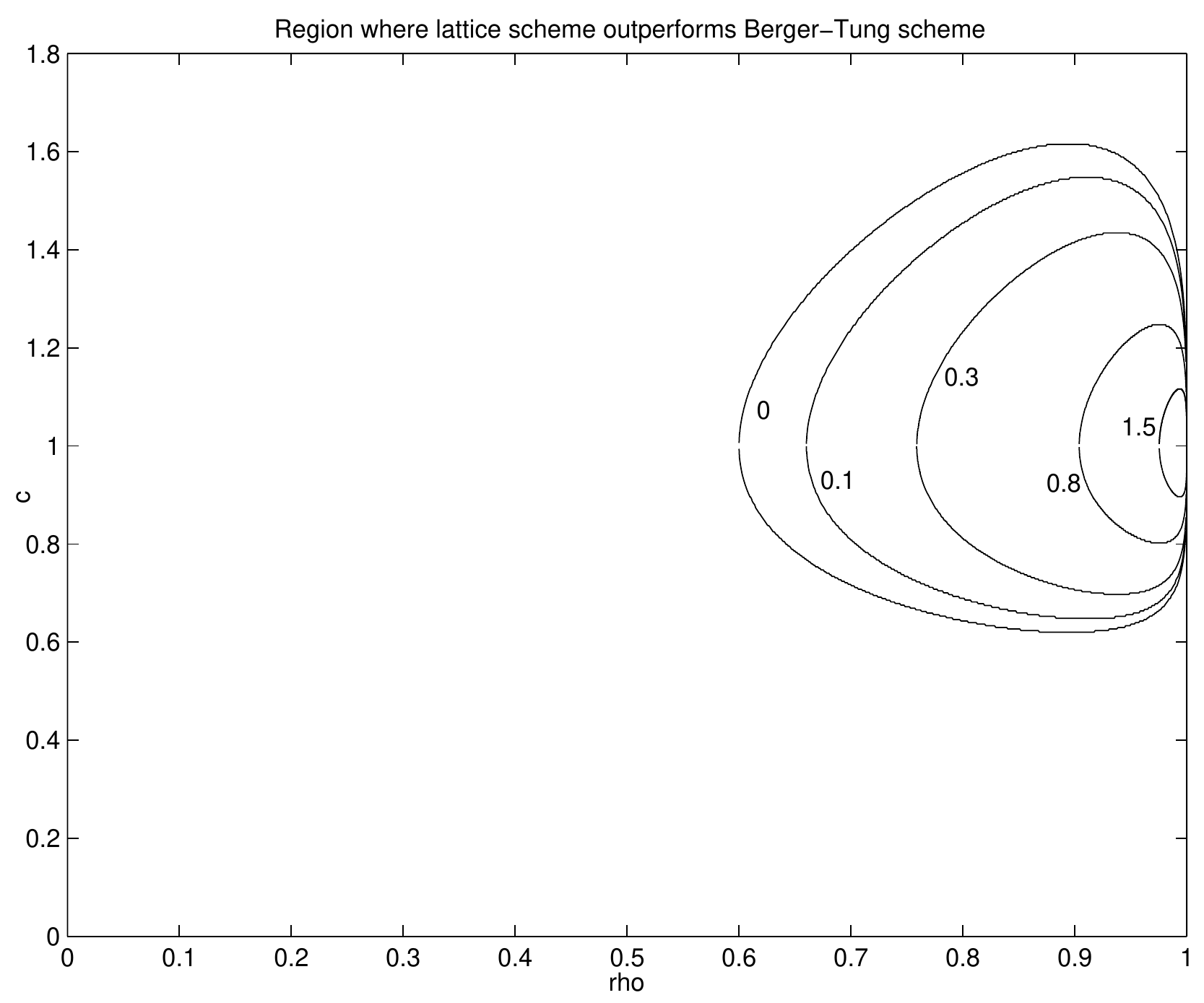}
\caption{Range of $(\rho,c)$ where the lattice scheme performs
better than the Berger Tung scheme} \label {fig:rhocrange}
\end{figure}

We observe that the lattice based scheme performs better than the
Berger-Tung based scheme for small distortions provided $\rho$ is
sufficiently high and $c$ lies in a certain interval. Fig.
\ref{fig:rhocrange} is a contour plot that illustrates this in
detail. The contour labeled $R$ encloses that region in which the
pair $(\rho,c)$ should lie for the lattice binning scheme to achieve
a sum rate that is at least $R$ units less than the sum rate of the
Berger-Tung scheme for some distortion $D$. Observe that we get
improvements only for $c>0$.

\section{Conclusion} \label{sec:conclusions}
We have thus demonstrated a lattice based coding scheme that
directly encodes the linear function that the decoder is interested
in instead of encoding the sources separately and estimating the
function at the decoder. For the case of two users, it is seen that
the lattice based coding scheme gives a lower sum-rate for certain
values of $\rho,c,D$. Hence, using a combination of the lattice
based and the Berger-Tung based coding schemes results in a better
rate-region than using any one scheme alone. For the case of
reconstructing a linear function of $K$ sources, we have extended
this concept to provide an inner bound to the optimal
rate-distortion function. Some parts of the inner bound are achieved
using a coding scheme that has the following structure: lattice
vector quantization followed by ``correlated'' lattice-structured
binning.

\section*{Acknowledgements} The authors would like to thank Dr.~Ram
Zamir and Dr.~Uri Erez of Tel Aviv University for helpful
discussions.

\appendix
\section{Derivation of Berger-Tung based scheme's sum rate}
\label{sec:btderive}

In this section, we derive the sum-rate of the Berger-Tung based
scheme given in equations
(\ref{eq:BTratecommon})-(\ref{eq:BTratecmore1}). The sum-rate of the
Berger-Tung based coding scheme is given by
\begin{align} \label{eq:appsumrate}
R_1+R_2 &\geq \frac{1}{2} \log \frac{(1+q_1)(1+q_2)-\rho^2}{q_1 q_2}
\end{align}
where $(q_1,q_2) \in \Bbb{R}^2_{+}$ should satisfy the distortion
constraint
\begin{align} \label{eq:appconstraint}
D &\geq \frac{q_1 \alpha + q_2 c^2 \alpha + q_1 q_2
\sigma_Z^2}{(1+q_1)(1+q_2)-\rho^2}
\end{align}
where $\mathbb{R}_{+}$ is the set of positive reals and $\alpha =
1-\rho^2$.

To minimize the sum-rate, we need to minimize the quantity given by
equation (\ref{eq:appsumrate}). Using the fact that the log function
is monotone and that $(q_1,q_2)$ must satisfy the distortion
constraint in equation (\ref{eq:appconstraint}), the minimization
problem is equivalent to minimizing
\begin{align}
\frac{(1+q_1)(1+q_2)-\rho^2}{q_1 q_2} &= \frac{q_1 \alpha + q_2 c^2
\alpha + q_1 q_2 \sigma_Z^2}{D q_1 q_2}
\end{align}
and this is equivalent to minimizing
\begin{align} \label{eq:appobjfunc}
\frac{1}{q_2} + \frac{c^2}{q_1}
\end{align}
subject to the constraint in equation (\ref{eq:appconstraint}).

Assuming that $(q_1,q_2)$ satisfy the distortion constraint with
equality, one can solve for $q_2$ in terms of $q_1$ to give
\begin{align}
q_2 &= \frac{\alpha D-q_1(\alpha-D)}{(c^2\alpha - D) +
q_1(\sigma_Z^2-D)}.
\end{align}
Substituting this in equation (\ref{eq:appobjfunc}) gives the
function to be minimized as a function of $q_1$ alone. The optimal
choice of $q_1$ is then
\begin{align}
q_1^{*} &= \text{argmin } \frac{q_1^2(\sigma_Z^2-D) + q_1 D(c^2-1) +
\alpha Dc^2}{-q_1^2(\alpha-D)+\alpha Dq_1}.
\end{align}
Differentiating with respect to $q_1$ and setting the derivative to
$0$ gives us a quadratic in $q_1$ whose roots are
\begin{align}
q_1^{*} &= \frac{\alpha c}{\rho-c} \,\, \text{ or }\,\, \frac{\alpha
cD}{2\alpha c-(\rho+c)D}
\end{align}
The second root given above is where the minima occurs. The $q_2$
value corresponding to this value of $q_1$ is
\begin{align}
q_2^{*} &= \frac{\alpha D}{2\alpha c^2- (1+\rho c)D}.
\end{align}
Note that these optimal values of $q_1$ and $q_2$ are positive only
for distortions in the range
\begin{align} \label{eq:appposDrange}
D &\leq \min \left\{ \frac{2\alpha c}{\rho+c} \, , \, \frac{2\alpha
c^2}{1+\rho c} \right\}.
\end{align}
For values of $D$ outside this range, the optimal strategy is to let
$q_1$ or $q_2$ go to $\infty$ which effectively means that we encode
and transmit only one source.

For $D$ in the range given in equation (\ref{eq:appposDrange}), the
sum rate $R_{\text{sum}} = R_1 + R_2$ is found by substituting
$q_1^{*}$ and $q_2^{*}$ in equation (\ref{eq:appsumrate}) to get
\begin{align} \label{eq:appsumrate1}
R_{\text{sum}} &\geq \frac{1}{2} \log \frac{4c(\alpha c- \rho
D)}{D^2} \quad D \leq \min \left\{ \frac{2\alpha c}{\rho+c} \, , \,
\frac{2\alpha c^2}{1+\rho c} \right\}.
\end{align}
For $D$ outside the range given in equation (\ref{eq:appposDrange}),
the minimum sum rate is attained by setting either $q_1$ or $q_2$ as
$\infty$. Which quantity goes to $\infty$ depends on which argument
of the min function in equation (\ref{eq:appposDrange}) is smaller;
equivalently on whether $c > 1$ or not. It is easy to see that if $c
< 1$, $q_2 = \infty$ and
\begin{align} \label{eq:appsumrate2}
R_{\text{sum}} &= \frac{1}{2} \log \frac {(1-\rho c)^2}{D-\alpha
c^2} \quad \text{for }D > \frac{2\alpha c^2}{1+\rho c},
\end{align}
and if $c>1$, $q_1 = \infty$ and
\begin{align} \label{eq:appsumrate3}
R_{\text{sum}} &= \frac{1}{2} \log \frac{(c-\rho)^2}{D-\alpha} \quad
\text{for }D > \frac{2\alpha c}{\rho+c}.
\end{align}
Combining equations (\ref{eq:appsumrate1}), (\ref{eq:appsumrate2})
and (\ref{eq:appsumrate3}) and taking the convex closure of the
resulting region, the complete rate region for the Berger-Tung based
scheme can be found.

\section{Existence of good nested lattices}
\label{sec:nestedlatticeproof}

In this section, we show the existence of nested lattices with any
finite degree of nesting such that all the lattice codes involved
are simultaneously good source and channel codes. More precisely, we
show the existence of a nested lattice $(\Lambda_1, \dots,
\Lambda_m)$, $\Lambda_m \subset \dots \subset \Lambda_1$ such that
$\Lambda_i$, $i = 1,\dots,m$ are simultaneously good source and
channel codes for sufficiently large lattice dimension $n$.

We use the same nested lattice ensemble as described in
\cite{zamirsnr}. For completeness sake, we include a description of
the ensemble.
\begin{itemize}
\item Start with a fixed $n$-dimensional lattice $\Lambda_2$ which is a good source and
channel code. The existence of such a lattice was shown in
\cite{zamirgood}. Let $G_{\Lambda_2}$ be the generator matrix of
$\Lambda_2$, i.e., $\Lambda_2 = G_{\Lambda_2} \cdot \Bbb{Z}^n$.
\item Construct a $k \times n$ matrix $G$ whose elements are drawn
according to an uniform i.i.d distribution over $\Bbb{Z}_p =
\{0,1,\dots,p-1\}$ where $p$ is an appropriately chosen prime.
\item Define the discrete codebook $\mathcal{C} = \{x \in
\Bbb{Z}_p^n \colon x = y \cdot G \text{ for some } y \in \Bbb{Z}_p^k
\}$.
\item Apply Loeliger's type A construction \cite{loeliger} to form
the lattice $\Lambda_1^{'} = p^{-1} \mathcal{C} + \Bbb{Z}^n$.
\item Transform $\Lambda_1^{'}$ to get the fine lattice $\Lambda_1
\triangleq G_{\Lambda_2} \cdot \Lambda_1^{'}$.
\item This construction of $\Lambda_1$ can be viewed equivalently as follows. From the
lattice $p^{-1} \Lambda_2$, pick $k < n$ points at random along with
all their multiples modulo-$\Lambda_2$. The resulting set of points
constitute the fine lattice $\Lambda_1$.
\end{itemize}
By construction, it follows that $\Lambda_2 \subset \Lambda_1$ with
the nesting ratio $\sqrt[n]{p^k}$. It was shown in \cite{zamirsnr}
that, with high probability, this construction will result in
$\Lambda_1$ being a good channel code. In \cite{dinesh-pradhan}, it
was shown that a nested lattice from this ensemble is, with high
probability, a good source code as well. By union bound, it then
follows that $\Lambda_1$ is simultaneously a good source and channel
code with high probability. It follows that there exists nested
lattices $(\Lambda_1, \Lambda_2)$ such that $\Lambda_2 \subset
\Lambda_1$ and both $\Lambda_1$ and $\Lambda_2$ are simultaneously
good source and channel codes.

By iterating this process (with $\Lambda_1$ playing the role of
$\Lambda_2$ in the construction above), one can obtain a nested
lattice code with any finite level of nesting. More precisely, for
any $m > 0$, one can show the existence of a nested lattice
$(\Lambda_1,\dots,\Lambda_m), \, \Lambda_m \subset \dots \subset
\Lambda_1$ such that all the lattices $\Lambda_i, i = 1,\dots,m$ are
simultaneously good source and channel codes. Moreover, this can be
accomplished for any choice of nesting ratios.

\section{Proof of convergence to Gaussianity of $e_q$}
\label{sec:eqconvproof}

In this section, we prove the claim that $e_q = e_{q_1}-e_{q_2}$
tends to a white Gaussian noise in the Kullback-Leibler divergence
sense. Note that $e_{q_1}$ and $e_{q_2}$ are independent.

We use the following properties of subtractive dither quantization
noise and the associated optimal lattice quantizers \cite{zamirlqn}.
\begin{itemize}
\item The subtractive dither quantization noise $e_{q_i}$ is uniformly
distributed over the basic Voronoi region $\mathcal{V}_{0,{1i}}$ of
the fine lattice $\Lambda_{1i}$ for $i=1,2$. It follows from
equation (\ref{eq:siglatdef}) that
\begin{align} \label{eq:appeqvar}
\mathbb{E} \parallel e_{q_i} \parallel^2 = n\sigma^2(\Lambda_{1i})
\quad \text{for }i=1,2.
\end{align}
\item For optimal lattice quantizers, the components of $e_{q_i}, i =
1,2$ are uncorrelated and have the same power,i.e., their
correlation matrices $\Sigma_{e_{q_i}}$ can be written as
\begin{align} \label{eq:covmat}
\Sigma_{e_{q_i}} = \sigma^2(\Lambda_{1i}) \mathbf{I}_{n \times n}
\quad \text{for }i=1,2.
\end{align}
\item For optimal lattice quantizers, as the lattice dimension $n
\rightarrow \infty$, the distribution of $e_{q_i}, i = 1,2$ tends to
a white Gaussian vector of same covariance in the Kullback-Leibler
divergence sense. Taking into account equation (\ref{eq:appeqvar}),
this can be written as
\begin{align}
\frac{1}{n} D\left(e_{q_i}
\parallel\mathcal{N}(0,\sigma^2(\Lambda_{1i})\mathbf{I}_{n \times n}) \right) \rightarrow 0 \quad
\text{for }i=1,2
\end{align}
in terms of the Kullback-Leibler divergence $D(. \parallel .)$ or
equivalently,
\begin{align} \label{eq:apphlim}
h(e_{q_i}) \rightarrow \frac{n}{2} \log 2\pi e
\sigma^2(\Lambda_{1i}) \quad \text{for }i=1,2
\end{align}
in terms of differential entropy $h(\cdot)$.
\end{itemize}

To show the convergence of $e_q$ to a white Gaussian random vector,
we use the entropy power inequality and the fact that for a given
covariance matrix, the Gaussian distribution maximizes differential
entropy.

The entropy power inequality \cite{cover-thomas} states that for two independent
$n$-dimensional random vectors $X$ and $Y$ (having densities),
\begin{align}
2^{\frac{2}{n}h(X+Y)} &\geq 2^{\frac{2}{n}h(X)}+2^{\frac{2}{n}h(Y)}.
\end{align}
This inequality applied to the subtractive dither quantization
noises gives
\begin{align} \label{eq:appepi}
2^{\frac{2}{n}h(e_{q_1} - e_{q_2})} &\geq 2^{\frac{2}{n}h(e_{q_1})}
+ 2^{\frac{2}{n}h(e_{q_2})}.
\end{align}
As $n \rightarrow \infty$, by equation (\ref{eq:apphlim}), the right
hand side of equation (\ref{eq:appepi}) tends to $2\pi e
(\sigma^2(\Lambda_{11})+ \sigma^2(\Lambda_{12}))$. So, we have the
following lower bound on the limit of the differential entropy of
$e_q$.
\begin{align} \label{eq:applowbound}
\lim_{n \rightarrow \infty} h(e_q) &\geq \frac{n}{2}\log 2\pi e
(\sigma^2(\Lambda_{11})+ \sigma^2(\Lambda_{12})).
\end{align}
To prove the inequality in the other direction, note that equation
(\ref{eq:covmat}) implies that the covariance matrix of $e_q$ is
$(\sigma^2(\Lambda_{11})+ \sigma^2(\Lambda_{12}))\mathbf{I}_{n
\times n}$. Since the Gaussian distribution maximizes differential
entropy for a given covariance matrix, we have
\begin{align} \label{eq:appupbound}
h(e_q) &\leq \frac{n}{2} \log 2 \pi e (\sigma^2(\Lambda_{11})+
\sigma^2(\Lambda_{12}))
\end{align}

Combining equations (\ref{eq:applowbound}) and
(\ref{eq:appupbound}), we have the desired result that (if optimal
lattice quantizers are used)
\begin{align}
\lim_{n \rightarrow \infty} h(e_q) &= \frac{n}{2} \log 2\pi
e(\sigma^2(\Lambda_{11})+ \sigma^2(\Lambda_{12})).
\end{align}
In words, $e_q$ tends in the Kullback-Leibler divergence sense to a
white Gaussian random vector with covariance matrix
$(\sigma^2(\Lambda_{11}) + \sigma^2(\Lambda_{12})) \mathbf{I}_{n
\times n}$.


\section{Derivation of optimal Lattice parameters}
\label{sec:optchoices} In the coding schemes of both Section
\ref{sec:defandresult} and Section \ref{sec:generalizations}, we
scale the sources before encoding them. Here, we briefly outline a
justification for the specific scaling constants used. We restrict
ourselves to the case where all the $K$ users encode their sources
using lattice binning. In the notation of Section
\ref{subsec:Kcase}, this corresponds to $\Theta = \{1,\ldots,K\}$.

Let the function to be reconstructed be $Z = \sum_{i=1}^{K} c_i X_i
= cX^n$. Here $c$ is a row vector with its $i$th component as $c_i$
and $X^n$ is a column vector of the sources $X_i$. $\Sigma$ is the
covaraince matrix of the random vector $X^n$. Let the $i$th encoder
scale its input by an arbitrary constant $\eta_i$. Let $\eta
\triangleq [\eta_1,\ldots,\eta_K]$. Choose a tuple $\mathcal{Q} =
(q_1,\ldots,q_K) \in \Bbb{R}_{+}^{K}$ just as in Section
\ref{subsec:latcodingKuser}.

It can be shown from analysis similar to the ones in Section
\ref{subsec:mainscheme} and \ref{subsec:Kcase} that the decoder can,
with high probability, reconstruct the function $\eta X^n + Q$ where
$Q$ approaches a white Gaussian noise of variance $q =
\sum_{i=1}^{K} q_i$. From equation (\ref{eq:Fdef}), it follows that
the function $f$ used for decoding is
\begin{align}
\hat{Z} &=  \left( \frac{c \Sigma \eta^T}{\eta \Sigma \eta^T + q }
\right) (\eta X^n + Q)
\end{align}
and the corresponding distortion is
\begin{align}
D &= \sigma_Z^2 - \frac{(c \Sigma \eta^T )^2}{\eta \Sigma \eta^T +
q}.
\end{align}
This fixes the value of $q$. The second moment of the channel code
used is $\sigma^2(\Lambda_2) = \text{Var}(\sum_i \eta_i X_i + q_i) =
\eta \Sigma \eta^T + q$. This gives us the rate tuple
\begin{align}
R_i &= \frac{1}{2} \log \frac{\eta \Sigma \eta^T + q}{q_i} \quad
\text{for } i = 1,\ldots,K
\end{align}
Eliminating $q_i$ using $q = \sum_i q_i$  gives us the rate region
\begin{align}
\sum_{i=1}^{K} 2^{-2R_i} &\leq 1 - (\sigma_Z^2-D) \frac{\eta \Sigma
\eta^T}{(c \Sigma \eta^T)^2}.
\end{align}
This rate region is largest when the RHS is maximum. Maximizing the
RHS as a function of $\eta$ results in $\eta = \xi \cdot c$ as the
only solution for some constant $\xi$. However, all constants $\xi$
result in the same rate region.


\begin{thebibliography}{10}

\bibitem{slepian73}
D.~Slepian and J.~K. Wolf, ``A coding theorem for multiple access channels with
  correlated sources,'' {\em Bell Syst. Tech. J.}, vol.~52, pp.~1037--1076,
  September 1973.

\bibitem{wyner74}
A.~D. Wyner, ``Recent results in {S}hannon theory,'' {\em IEEE Trans. on Inform.
  Theory}, vol.~20, pp.~2--10, January 1974.

\bibitem{wyner75}
A.~D. Wyner, ``On source coding with side information at the decoder,'' {\em
  IEEE Trans. on Inform. Theory}, vol.~IT-21, pp.~294--300, May 1975.

\bibitem{ahlswede-korner}
R.~Ahlswede and J.~Korner, ``Source coding with side information and a converse
  for degraded broadcast channels,'' {\em IEEE Trans. Inform. Theory}, vol.~IT-
  21, pp.~629--637, November 1975.

\bibitem{wynerziv}
A.~D. Wyner and J.~Ziv, ``The rate-distortion function for source coding with
  side information at the decoder,'' {\em IEEE Trans. Inform. Theory}, vol.~IT-
  22, pp.~1--10, January 1976.

\bibitem{berger77}
T.~Berger, ``{Multiterminal source coding},'' in {\em Lectures presented at
  {CISM} summer school on the Inform. Theory approach to communications}, July
  1977.

\bibitem{bergertung}
S.-Y. Tung, {\em Multiterminal source coding}.
\newblock PhD thesis, School of Electrical Engineering, Cornell University,
  Ithaca, NY, May 1978.

\bibitem{berger-housewright}
T.~Berger, K.~B. Housewright, J.~K. Omura, S.~Tung, and J.~Wolfowitz, ``An
  upper bound on the rate distortion function for source coding with partial
  side information at the decoder,'' {\em IEEE Trans. Inform. Theory}, vol.~IT-
  25, pp.~664--666, November 1979.

\bibitem{gelfandpinsker}
S.~Gelfand and M.~Pinsker, ``Coding of sources on the basis of observations
  with incomplete information,'' {\em Problemy Peredachi Informatsii}, vol.~15,
  pp.~45--57, Apr-June 1979.

\bibitem{kornermarton}
J.~Korner and K.~Marton, ``How to encode the modulo-two sum of binary
  sources,'' {\em IEEE Trans. Inform. Theory}, vol.~IT-25, pp.~219--221, March
  1979.

\bibitem{han79}
T.~S.~Han, ``Source coding with cross observations at the encoders,''
{\em IEEE Trans. Inform. Theory}, vol.~IT-25, pp.~360--361, May
  1979.


\bibitem{csiszar}
I.~Csisz\'{a}r and J.~Korner, {\em Information Theory: Coding Theorems for
  Discrete Memoryless Systems}.
\newblock Academic Press Inc. Ltd., 1981.

\bibitem{cover-thomas}
T.~M. Cover and J.~A. Thomas, \emph{Elements of Information Theory}.\hskip 1em
  plus 0.5em minus 0.4em\relax New York:Wiley, 1991.


\bibitem{han-kobayashi}
T.~S. Han and K.~Kobayashi, ``A dichotomy of functions {F(X,Y)} of correlated
  sources {(X,Y)},'' {\em IEEE Trans. on Inform. Theory}, vol.~33, pp.~69--76,
  January 1987.

\bibitem{ahlswede-han}
R.~Ahlswede and T.~S. Han, ``On source coding with side information via a
  multiple-access channel and related problems in multi-user information
  theory,'' {\em IEEE Trans. on Inform. Theory}, vol.~29, pp.~396--412, May
  1983.

\bibitem{berger-yeung}
R.~W. Yeung and T.~Berger, ``Multiterminal source coding with one distortion
  criterion,'' {\em IEEE Trans. on Inform. Theory}, vol.~35, pp.~228--236,
  March 1989.

\bibitem{viswanathan-berger-old}
H.~Viswanathan, Z.~Zhang, and T.~Berger, ``The {CEO} problem,'' {\em IEEE
  Trans. on Inform. Theory}, vol.~42, pp.~887--902, May 1996.

\bibitem{viswanathan-berger-new}
H.~Viswanathan and T.~Berger, ``The quadratic {G}aussian {CEO} problem,'' {\em
  IEEE Trans. on Inform. Theory}, vol.~43, pp.~1549--1559, September 1997.

\bibitem{viswanath}
P.~Viswanath, ``{Sumrate of multiterminal {G}aussian source coding },'' in {\em
  DIMACS workshop on Network Information Theory}, (Piscataway, NJ), April 2002.

\bibitem{yamamoto-itoh}
H.~Yamamoto and K.~Itoh, ``Source coding theory for multiterminal communication
  systems with a remote source,'' {\em The Transactions of the IECE of Japan},
  vol.~E-63, pp.~700--706, October 1980.

\bibitem{flynn-gray}
T.~J. Flynn and R.~M. Gray, ``Encoding of correlated observations,'' {\em IEEE
  Trans. Inform. Theory}, vol.~IT-33, pp.~773--787, November 1987.

\bibitem{dobrushin}
R.~Dobrushin and B.~Tsybakov, ``Information transmission with additional
  noise,'' {\em IRE Trans. Inform. Theory}, vol.~IT- 18, pp.~S293--S304, 1962.

\bibitem{witsenhausen}
H.~S. Witsenhausen, ``Indirect rate distortion problems,'' {\em IEEE Trans.
  Inform. Theory}, vol.~IT- 26, pp.~518--521, September 1980.


\bibitem{oohama97}
Y.~Oohama, ``Gaussian multiterminal source coding,'' {\em IEEE Trans. Inform.
  Theory}, vol.~IT- 43, pp.~1912--1923, November 1997.

\bibitem{oohama98}
Y.~Oohama, ``The rate-distortion function for the quadratic {G}aussian {CEO}
  problem,'' {\em IEEE Trans. Inform. Theory}, vol.~IT-44, pp.~1057--1070, May
  1998.

\bibitem{oohama05}
Y.~Oohama, ``Rate-distortion theory for {G}aussian multiterminal source coding
  systems with several side informations at the decoder,'' {\em IEEE Trans.
  Inform. Theory}, vol.~IT-51, pp.~2577--2593, July 2005.


\bibitem{oohama_allerton05}
Y.~Oohama, ``{Rate distortion region for separate coding of correlated
  {G}aussian remote observations},'' in {\em Allerton Conference}, (Monticello,
  IL), September 2005.

\bibitem{oohama_ita06}
Y.~Oohama, ``{Separate source coding of correlated {G}aussian remote
  sources},'' in {\em Workshop on Information theory and Applications (ITA)},
  (San Diego, CA), January 2006.

\bibitem{wagner-anantharam}
A.~B. Wagner and V.~Anantharam, ``{An improved outer bound for the
  multiterminal source-coding problem},'' in {\em IEEE International Symposium
  on Information Theory (ISIT '05)}, (Adelaide, Australia), September 2005.

\bibitem{prabhakaran-ramchandran-tse}
V.~Prabhakaran, K.~Ramchandran, and D.~Tse, ``{Rate region of the quadratic
  {G}aussian {CEO} problem},'' in {\em IEEE International Symposium on
  Information Theory (ISIT '04)}, (Chicago, IL), p.~117, June 2004.

\bibitem{orlitsky-roche}
A.~Orlitsky and J.~R. Roche, ``Coding for computing,'' {\em IEEE Trans. Inform.
  Theory}, vol.~IT-47, pp.~903--917, March 2001.

\bibitem{yamamoto}
H.~Yamamoto, ``Wyner-{Z}iv theory for a general function of the correlated
  sources,'' {\em IEEE Trans. Inform. Theory}, vol.~IT-28, pp.~803--807,
  September 1982.

\bibitem{feng-effros-savari}
H.~Feng, M.~Effros, and S.~A. Savari, ``{Functional source coding for networks
  with receiver side information},'' in {\em Proc. of the 42nd annual Allerton
  conference on Communication, Control and Computing}, (Monticello, IL),
  September 2004.

\bibitem{gastpar}
M.~Gastpar, ``The {W}yner-{Z}iv problem with multiple sources,'' {\em IEEE Trans.
  Inform. Theory}, vol.~IT-50, pp.~2762--2768, November 2004.


\bibitem{wagner-tavildar-viswanath-old}
A.~B. Wagner, S.~Tavildar, and P.~Viswanath, ``The rate-region of the quadratic
  {G}aussian two-terminal source-coding problem,'' {\em \tt arXiv:cs.IT/0510095}.

\bibitem{wagner-tavildar-viswanath-new}
S.~Tavildar, P.~Viswanath, and A.~B. Wagner, ``{The {G}aussian many-help-one
  distributed source coding problem},'' in {\em Proc. of the 2006 IEEE Inform.
  Theory Workshop (ITW '06)}, (Chengdu, China), pp.~596--600, October 2006.

\bibitem{jana-blahut}
S.~Jana and R.~Blahut, ``Achievable region for multiterminal source
coding with lossless decoding in all sources except one,'' to appear
in {\em IEEE Inform. Theory Workshop (ITW '07)}, (Lake Tahoe, CA),
September 2007.

\bibitem{jana1}
S.~Jana, ``Unified theory of source coding: Part I - two terminal
problems,'' {\em \tt arXiv:cs/0508118v2}.

\bibitem{jana2}
S.~Jana, ``Unified theory of source coding: Part II - multiterminal
problems,'' {\em \tt arXiv:cs/0508119v1}


\bibitem{csiszar82}
I.~Csiszar, ``Linear codes for sources and source networks: Error exponents,
  universal coding,'' {\em IEEE Trans. Inform. Theory}, vol.~IT- 28,
  pp.~585--592, July 1982.


\bibitem{zamir-feder}
R.~Zamir and M.~Feder, ``On universal quantization by randomized uniform
  lattice quantizers,'' {\em IEEE Trans. Inform. Theory}, vol.~IT- 38,
  pp.~428--436, March 1992.

\bibitem{zamirlqn}
R.~Zamir and M.~Feder, ``On lattice quantization noise,'' {\em IEEE Trans.
  Inform. Theory}, vol.~IT-42, pp.~1152--1159, July 1996.

\bibitem{shamai-verdu-zamir}
S.~Shamai, S.~Verdu, and R.~Zamir, ``Systematic lossy source/channel coding,''
  {\em IEEE Trans. on Inform. Theory}, vol.~44, pp.~564--579, March 1998.

\bibitem{zamir-berger}
R.~Zamir and T.~Berger, ``Multiterminal source coding with high resolution,''
  {\em IEEE Trans. Inform. Theory}, vol.~IT-45, pp.~106--117, January 1999.

\bibitem{zamirmulti}
R.~Zamir, S.~Shamai, and U.~Erez, ``Nested linear/lattice codes for structured
  multiterminal binning,'' {\em IEEE Trans. Inform. Theory}, vol.~IT-48,
  pp.~1250--1276, June 2002.

\bibitem{zamirsnr}
U.~Erez and R.~Zamir, ``Achieving 1/2 log(1+{SNR}) on the {AWGN} channel with
  lattice encoding and decoding,'' {\em IEEE Trans. Inform. Theory}, vol.~IT-
  50, pp.~2293--2314, October 2004.

\bibitem{zamirgood}
U.~Erez, S.~Litsyn, and R.~Zamir, ``Lattices which are good for (almost)
  everything,'' {\em IEEE Trans. Inform. Theory}, vol.~IT- 51, pp.~3401--3416,
  October 2005.

\bibitem{minkowski_1904}
H.~Minkowski, ``Dichteste gitterf\"{o}rmige Lagerung kongruenter
K\"{o}rper,'' {\em Nachr. Ges. Wiss. G\"{o}ttingen}, pp.~311--355,
1904.

\bibitem{hlawka}
E.~Hlawka, ``Zur Geometrie der Zahlen,'' {\em Math.Z.}, vol.~49,
pp.~285--312, 1944.

\bibitem{kershner_1939}
R.~Kershner, ``The number of circles covering a set,'' {\em Amer.
Jour. Math.}, vol.~61, pp.~665--671, 1939.

\bibitem{rogers_book}
C.~A. Rogers, {\em Packing and Covering}. \newblock Cambridge
University Press, Cambridge, 1964.

\bibitem{loeliger}
H.~A. Loeliger, ``Averaging bounds for lattices and linear codes,''
{\em IEEE
  Trans. Inform. Theory}, vol.~IT- 43, pp.~1767--1773, November 1997.

\bibitem{dinesh-pradhan}
D.~Krithivasan and S.S.~Pradhan, ``A proof of the existence of good
nested lattices,'' {\em \tt
http://www.eecs.umich.edu/techreports/systems/cspl/cspl-384.pdf}.

\bibitem{conway-sloane}
J.~H.~Conway and N.~J.~A.~Sloane, {\em Sphere Packings, Lattices and
Groups}.
\newblock Springer, 1992.

\bibitem{kirac-vaidyanathan}
A.~Kirac and P.~Vaidyanathan, ``Results on lattice vector quantization with
  dithering,'' {\em IEEE Trans. Circuits and Systems II: Analog and Digital
  Signal Processing}, vol.~43, pp.~811--826, December 1996.

\bibitem{vaishampayan-sloane-servetto}
V.~A.~Vaishampayan, N.~J.~A.~Sloane and S.~D.~Servetto,
``Multiple-description vector quantization with lattice codebooks:
design and analysis,'' {\em IEEE Trans. Inform. Theory}, vol.~IT-47,
pp.~1718--1734, July 2001.

\bibitem{dayan-zamir}
Y.~Frank-Dayan and R.~Zamir, ``Dithered lattice-based quantizers for
multiple descriptions,'' {\em IEEE Trans. Inform. Theory},
vol.~IT-48, pp.~192--204, January 2002.

\bibitem{goyal-kelner}
V.~K.~Goyal, J.~A.~Kelner and J.~Kovacevic, ``Multiple description
vector quantization with a coarse lattice,'' {\em IEEE Trans.
Inform. Theory}, vol.~IT-48, pp.~781--788, March 2002.

\bibitem{diggavi-sloane-vaishampayan}
S.~N.~Diggavi, N.~J.~A.~Sloane and V.~A.~Vaishampayan, ``Asymmetric
multiple description lattice vector quantizers,'' {\em IEEE Trans.
Inform. Theory}, vol.~IT-48, pp.~174--191, January 2002.

\bibitem{ostergaard}
J.~Ostergaard, {\em Multiple-description lattice vector
quantization}.
\newblock PhD thesis, Delft University of Technology, Netherlands,
June 2007.

\bibitem{Poltyrev}
G.~Poltyrev, ``On coding without restrictions for the {AWGN} channel,'' {\em
  IEEE Trans. on Inform. Theory}, vol.~40, pp.~409--417, March 1994.


\end{thebibliography}
\end{document}